\documentclass[aps,pra,showpacs]{revtex4}
\usepackage{graphics,epsfig}
\begin{document}
\title{Surface modes of  ultra-cold atomic clouds 
with  very large number
of vortices}
\author{M. A. Cazalilla}
\affiliation{The Abdus Salam ICTP,
Strada Costiera 11, 34014 Trieste, Italy, }
\affiliation{ Donostia International Physics Center (DIPC),
Manuel de Lardizabal 4, 20018 Donostia, Spain.}
\begin{abstract}
We study the surface modes of some of the vortex liquids recently found
by means of exact diagonalizations in systems of rapidly rotating bosons.
In contrast to the surface modes of Bose condensates, we find that 
the surface waves have a frequency linear in the excitation angular
momentum, $\hbar l > 0$. Furthermore,  in analogy with the edge waves of
electronic quantum Hall states, these excitations are {\it chiral}, that is, 
they can be excited only for values of  $l$ that increase 
the total angular momentum of the vortex liquid. However,  differently 
from the quantum Hall phenomena for electrons, we also find other
excitations that are approximately degenerate in the laboratory frame with the surface modes, 
and which decrease  the total angular momentum by $l$ quanta.
The surface modes of the Laughlin, as well as other scalar and vector boson states
are analyzed, and their {\it observable} properties characterized.
We argue that measurement of the response of a vortex liquid to
a weak time-dependent  potential that imparts angular momentum to the system should 
provide valuable information to characterize the vortex liquid. In particular,
the intensity of the signal of the surface waves in the dynamic
structure factor has been studied and found to depend on the type
of vortex liquid. We  point out that the existence of 
surface modes has observable consequences on the
density profile of the Laughlin state. These features are due
to the strongly correlated behavior of  atoms in the vortex liquids. 
We point out that these correlations should be  responsible for
a remarkable stability of  some  vortex  liquids  with respect to three-body 
losses.
\end{abstract}
\maketitle

\section{INTRODUCTION}

 Everybody is familiar with the phenomenon of vortices. 
We observe it every time we watch water down a drain. 
In classical fluids,  it  is a consequence
of  broken Galilean symmetry~\cite{SSB}: A fluid that fills a region of
space provides us with a privileged reference frame, namely the one
where the fluid is at rest.  Nature tries to restore the full Galilean invariance
at the vortex core, which explains why the fluid density 
drops to zero (or almost zero) there.
In   superfluids, however, vortices exhibit certain peculiarities 
not seen in classical fluids. This is because quantum 
coherence is maintained throughout the superfluid
volume,  and  since the wave function must be singly valued   
the circulation around a vortex is quantized. 
Therefore, the stability of vortices is ensured by
{\it topological} rather than {\it dynamical} reasons~\cite{vorticity}.  
Nevertheless, a vortex carrying $n>1$
circulation quanta is unstable with respect to 
decay into $n$  singly quantized vortices. When many
of these have appeared, they 
form a triangular lattice, the Abrikosov lattice. In ultra-cold 
atomic gases the formation of this lattice has been
observed in a recent experiment~\cite{experiments}. 
If the rotation frequency, $\Omega$, is
further increased to approach the trap frequency~\cite{trap}, $\omega_{\perp}$, 
several authors~\cite{Cooper01, Sinova02,Gosh02} have pointed to the 
interesting possibility that the Abrikosov lattice melts
due to quantum fluctuations.  This regime  
($\Omega \lesssim \omega_{\perp}$) is known as the {\it critical
rotation} limit,  and some experiments have already begun 
to be explore it~\cite{Rosenbusch02}.

What kind of phenomena can be expected to emerge in the critical
rotation limit? Under certain conditions, which will be discussed
below, it has been predicted that bosons organize themselves 
in  highly correlated, two-dimensional,   ``vortex liquids''. 
These are states that cannot be described by the
standard (i.e. Gross-Pitaevskii) mean field theory~\cite{Cooper99, Cooper01}.
Instead, they seem to be quite accurately described by
microscopic wave functions~\cite{Cooper99,Cooper01}, 
closely related to those used for electrons
in the context of the fractional quantum Hall effect 
(FQHE)~\cite{Laughlin83, Yoshioka02}. 
Recent numerical studies 
using periodic boundary conditions~\cite{Cooper01}
have shown that, just like their FQHE counterparts, 
homogeneous vortex liquids are incompressible. This means that changing 
their density requires a finite amount of energy, 
signaling the existence of a spectral gap.
When carried out for small systems in a harmonic trap, 
exact diagonalization studies~\cite{Cooper99, Wilkin00} 
found a series  particularly stable states at ``magic values'' of
the angular momentum, some of them related to the 
homogeneous vortex liquids~\cite{Regnault02}.
\begin{figure}[ht]
\centerline{\resizebox{5cm}{!}{\includegraphics{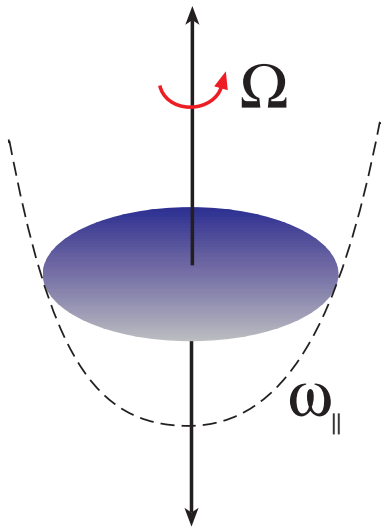}}}
\caption{Rapidly rotating cloud of ultra-cold atoms in the critical 
rotation regime $\Omega \lesssim \omega_{\perp}$.}
\label{fig0}
\end{figure}

 In this work we address the following question:
Which are the experimental signatures of these vortex 
liquids? The question is relevant since
experiments driving ultra-cold atom clouds to the critical rotation
limit are likely to proliferate in the near future.
It is also important to find ways to characterize
the vortex liquids by {\it non-destructive} means, especially
if one intends to use their entanglement properties
for quantum computing purposes~\cite{Kitaev97}.
Below we shall try to answer the previous question 
by studying the properties of  surface excitations
of some of these states. Since the vortex liquids
are effectively  two dimensional,  by ``surface'' we mean
the {\it one-dimensional boundary} in the plane perpendicular to the
rotation axis. Excitations along the rotation axis will
occur at  much higher energies and  are therefore  frozen 
at the low temperatures of interest here. As we shall 
argue in section~\ref{sectlaughlin},  deformations of the boundary 
correspond to the  lowest energy excitations of the system.
Their spectroscopic analysis should provide valuable
information on what kind of vortex liquid one is
dealing with.  In the case of the FQHE exhibited by
two-dimensional electron gases under  strong magnetic
fields, characterization of the state of the sample is  usually achieved
by means of transport measurements. In particular, a measurement 
of the Hall conductance
automatically yields the filling fraction $\nu$, which in the
present context and in the limit of large particle and vortex 
numbers can be defined
as the ratio of the number of particles to the number
of vortices, i.e. $\nu = N/N_v$. However, nothing like 
transport spectroscopy is yet  available for 
ultra-cold atomic clouds. 
Instead, experiments where the
trapping potential is weakly deformed in a time-dependent 
fashion, therefore exciting surface modes,  seem more feasible.  
That this is feasible with a {\it rotating} cloud containing a large number of vortices
has  been already demonstrated in a recent experiment~\cite{Engels02}  
(however,  not yet in the regime in which we are interested here).

The study of the surface modes  is also interesting because, at 
low temperatures, 
these excitations should dominate the properties of the system that
can be probed by weakly coupling to it.
Furthermore, it has been shown in the 
context of the FQHE~\cite{Wen92,Wen95} that  analyzing 
the boundary excitations is  the way to characterize 
the  bulk states  (i.e. their {\it quantum orders}~\cite{Wen95}).
What is more, the boundary of a vortex liquid is also 
a clean  example of an exotic type  of quantum liquid: The {\it chiral}
Luttinger liquid. The main difference with the  {\it electronic} FQHE is
that the constituent particles of the system are {\it neutral}
bosonic atoms, which interact via short-range interactions.
This setup offers some advantages over the 
two-dimensional  electron gas at high magnetic field. 
The systems are intrinsically clean, and the absence 
of long-range interactions does not lead to some complications 
introduced by the Coulomb potential 
(e.g. edge reconstructions~\cite{Yoshioka02}).
One can also consider atoms with internal degrees of freedom,
which opens the possibility to  study novel quantum Hall states~\cite{Paredes02,Reijnders02}. 
The disadvantages are  the fragility of the vortex liquids, which will
require careful experiments with  small numbers of particles, $N \lesssim 100$ 
(although many replicas of a rotating cloud can 
be created in an optical lattice~\cite{Ho02}). In the future,
other methods to ``simulate'' high magnetic fields using 
optical lattices  may become available~\cite{Jaksch02}, and this could 
lift the constraint on the particle number. However, large systems
confined by smooth potentials, like the ones used to trap ultra-cold
atoms, should exhibit more complicated 
boundary structures like ``composite edges''~\cite{Wen92}, 
with many branches of surface modes. Nevertheless,
the theory developed here should provide
the basis for the understanding of such systems.

	The experimental achievement of 
the vortex liquid states to be discussed below 
is challenging, but on the theoretical side 
a complete understanding of their properties
also poses many challenges. The existing work 
has  mainly focused on  ground states
of scalar~\cite{Cooper99,Cooper01} 
and higher-spin bosons~\cite{Paredes01,Ho02,Reijnders02},
but the situation is far from being as clear as in
the FQHE for electrons~\cite{Yoshioka02}.
In this paper, we shall consider the surface  
excitations of several  ground states of
scalar and vector bosons. One of them, the Laughlin state, 
is an exact ground state for rotating  scalar bosons. 
The other ground states for scalar
bosons that we will consider are approximate, 
and have been found to exhibit  good
overlap with the exact ground states obtained by
exact diagonalization methods~\cite{Cooper99,Wilkin00,Cooper01,Regnault02}.
We shall consider the Laughlin state first, for which
the microscopic construction of the surface waves
will be presented. An effective field theory will
be subsequently developed and shown to
agree with the microscopic theory.
Although the assumptions for the effective theory 
strictly hold in the large $N$ limit,  it is well established numerically 
(e.g. ~\cite{Lee91,Wen92,Mandal02})  
that it  also applies  to small systems with $N \sim 10$,
within the experimental reach in the near future.
 
  As many of the concepts and methods introduced and used in this paper
should be new to the community working in ultra-cold atom systems, we have adopted
a pedagogical approach at the cost of producing a longer paper. 
We have also tried to make the article as self-contained as
possible, but without omitting references to the original literature
on the quantum Hall effect, which should be consulted for 
more extended explanations. 
The paper is organized as follows: In the next section, the
experimental conditions for the existence of vortex liquids are discussed. 
Next, the simplest (from the theoretical point of view) type of vortex liquid, namely the
Laughlin liquid, is considered. In section~\ref{other} surface excitations of more 
complicated (but approximate)  states of scalar bosons are studied. In the following section, 
we take up vector bosons and  analyze the surface excitations 
of their  singlet states. Finally, the conclusions of this work 
can be found in Sect.~\ref{concl}. In the appendices we include
some supplementary material. Thus appendix~\ref{appa} 
discusses the relation between the laboratory and rotating frames,
and how measurements in the laboratory frame are related to
calculations in the rotating frame. We also discuss the generalized
Kohn's theorem in this appendix, and how it constrains the energy and
peak intensity of the dipole mode. Appendix~\ref{appb}, however, gives a 
brief introduction to 
the plasma formalism for quantum Hall wave functions. More importantly,
we also show  there how to obtain the wave functions from correlation functions of 
the field theory that describes the boundary excitations.
	
\section{Experimental conditions for the observation of vortex liquids}
\label{disgr}

 Before proceeding any further, it will be useful  to recall the conditions
under which it is expected  that a rotating cloud of ultra-cold 
atoms will become a vortex liquid.
The strong analogy between the quantum mechanics of rotating particles 
in a harmonic well and charged particles moving in two-dimensions
under a strong magnetic field will be useful here; therefore, 
the appearance of Landau levels is expected. To see this,  notice  that
the one-body part of the Hamiltonian in the {\it rotating} reference frame (see 
Appendix~\ref{appa}),
\begin{equation}\label{eq1}
H_j =   \frac{{\bf p}_j^2 + 
p^2_{zj}}{2M}  + 
\frac{M}{2} \left( \omega^2_{\perp} {\bf r}^2_j + \omega^2_{||} z^2_j \right)
- \Omega \: \hat{\bf z} \cdot {\bf L}_j  
\end{equation}
can be written in two equivalent ways:
\begin{equation}\label{eq2}
 H^{(1)}_j = \frac{({\bf p}_j - M \Omega\: \hat{\bf z}\times {\bf r}_j)^2 }{2M} + 
\frac{M}{2}(\omega^2_{\perp} - \Omega^2)\,  {\bf r}^2_j +
\frac{p^2_{zj}}{2M} +  \frac{M \omega^2_{||}}{2} z^2_j,
\end{equation}
and
\begin{equation}\label{eq3}
 H^{(2)}_j = \frac{({\bf p}_j - M \omega_{\perp}\: \hat{\bf z} \times {\bf r}_j)^2}{2M} 
+(\omega_{\perp}-  \Omega)\,  \hat{\bf z}\cdot {\bf L}_j +
\frac{p^2_{zj}}{2M} + \frac{M \omega^2_{||}}{2} z^2_j.
\end{equation}
We have assumed an axially symmetric trap with the 
rotation axis coinciding with the
symmetry axis of the trap; the following notation has 
been introduced: ${\bf r}_j = (x_j, y_j)$
and ${\bf p}_j = (p_{xj}, p_{yj})$ are the $j$-th particle position and momentum,
respectively, while ${\bf L}_j = {\bf r}_j \times {\bf p}_j$ is the angular momentum
along the rotation axis. In both forms of $H_j$, it is manifest that particles, in the rotating
reference frame, feel an {\it effective} magnetic field directed along $\bf \hat{z}$, 
whose magnitude is proportional to either $M \Omega$  or $M \omega_{\perp}$, 
depending on the form,  $M$ being the atom mass.  
Perhaps the first form,  $H^{(1)}_j$,   shows more explicitly 
what happens as $\Omega$ approaches $\omega_{\perp}$,
and the system enters the critical rotation regime: First,
notice that particles under 
rotation experience a reduced confinement in 
the plane perpendicular to the rotation axis. This 
means that, as $\Omega$ is increased towards the 
trap frequency $\omega_{\perp}$, more and more 
particles will accommodate into the lowest axial
level  (this can be helped by making $\omega_{||} > \omega_{\perp}$~\cite{trap}). 
Eventually,  when all the particles are in this level,
the system  effectively becomes two-dimensional. 
At the same time, increasing
$\Omega$ increases the strength of the 
effective magnetic field and therefore, for
sufficiently weak interactions, all particles 
will make its way into the lowest Landau
level (LLL).

\begin{figure}[t]
\centerline{\resizebox{6cm}{!}{\includegraphics{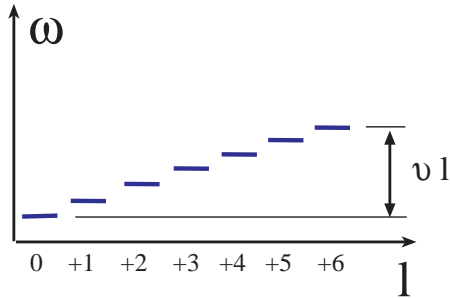}}}
\caption{For non-interacting particles, the large degeneracy of the lowest
Landau level  is lifted only by the confining potential. In the reference frame that rotates
with the system,  a ``band'' of width  $\approx \hbar\upsilon N/\nu$ is formed.}
\label{fig3}
\end{figure}
 To study the quasi two-dimensional system in the critical rotation regime 
with all the particles lying in the lowest Landau and axial levels,  it is more convenient
to use the second form of the Hamiltonian, $H^{(2)}_j$. Dropping the axial
part and introducing the  following operators: 
\begin{eqnarray}
\pi_j &=& \pi_{jx} + i \pi_{jy}, \\
\bar{\pi}_j &=& \pi_{jx} - i \pi_{jy}, \label{pibar} 
\end{eqnarray}
where $\pi_{jx} = p_{jx} + M\omega_{\perp}y_j$ and $\pi_{jy} = p_{jy} - M\omega_{\perp}x_j$,
and $\left[\pi_{j}, \bar{\pi}_{j} \right] = 4M \hbar \omega_{\perp}$,  allows us to write:
\begin{equation}
H^{(2)}_{j} = \frac{\bar{\pi}_j\pi_j}{2M} + (\omega_{\perp} - \Omega) L_j + \hbar \omega_{\perp}.
\end{equation}
One can easily convince oneself that this form is  diagonal after noticing that 
$\pi_j$ and $\bar{\pi}_j$ are nothing but ladder operators, which move particles between consecutive
Landau levels: $\bar{\pi}_j$ promotes the $j$-th particle to a higher Landau level whereas
$\pi_j$ undoes this operation. Thus, if a particle is in the LLL, the corresponding $\pi_j$ must
annihilate the state. Once  diagonalized, one finds  that the  single-particle orbitals in the LLL 
have energies (in the rotating frame) equal to $\epsilon(l) = \hbar \omega_{\perp} 
+ \hbar \upsilon l$, where $\upsilon \equiv   (\omega_{\perp} - \Omega) \ll \omega_{\perp}$
and $\hbar l$ is the angular momentum. They form a ``band'' (see Fig.~\ref{fig3})
within the LLL. The orbitals themselves take the form:
\begin{equation}\label{orbital}
\varphi_l(x,y) = \frac{1}{\ell \sqrt{\pi l!}} \left(\frac{z}{\ell}\right)^l \: e^{-|z|^2/2\ell^2},
\end{equation}
where $\ell = \sqrt{\hbar/M\omega_{\perp}}$ is the 
oscillator length of the trap, 
and $z = x+ i y$ denotes the position of the 
particle on the plane perpendicular to
the rotation axis. As the system is effectively
two dimensional,  we have omitted the 
orbital describing the motion perpendicular to the plane.

 We are now ready to take up the discussion 
of the conditions under which the scenario
described above can be experimentally realized. Interactions
between the particles are needed for the stability of the vortex liquids
to be discussed below. Taking them into account,  
the total Hamiltonian reads $H = \sum_{j=1}^{N}  H_j +
\sum_{i<j=1}^{N} V_{ij}$,  $V_{ij} = 
g \, \delta^{(2)}({\bf r}_{i} - {\bf r}_{j}) \delta(z_i-z_j)$,
where $g$ is related to the s-wave scattering length, $a$, as usual: $g = 4\pi\hbar^2a/M$.
The interactions should be neither too weak nor too
strong. Since $\hbar \upsilon$ is the level spacing 
of the single-particle band in the LLL, we must have that
$gn \gg \hbar \upsilon$ but $gn < N\hbar\upsilon/\nu$, where
$n$ is the mean density of the system.
The last condition is the requirement that the width of 
single-particle band must be larger than 
typical interaction energy, $gn$. The filling of the
band is determined by the total angular momentum
through the {\it filling fraction} $\nu \approx N^2/2L$.
Hence $N/\nu \approx 2L/N$ measures the mean
angular momentum per particle. Finally, since all the particles must
be in the LLL, mixing with higher Landau levels is avoided
provided that $ N\hbar\upsilon/\nu \ll \hbar\Omega$
and $gn \ll \hbar\Omega$ and at temperatures $T \ll \hbar \Omega$.
Although  these conditions look 
very demanding, we expect the rapid progress that has 
characterized the field over the last years  will make vortex 
liquids experimentally available  in a near future.

%
%
\section{Surface modes of the bosonic Laughlin state}\label{sectlaughlin}
 We begin by considering
the simplest case, from the theoretical point of view. A droplet of $N$ 
bosons in the  Laughlin state is described by the following wave function
\begin{equation}\label{laughlin}
\Phi_{m}(z_1,\ldots, z_N) = \prod_{i<j=1}^{N} \left( z_i - z_j \right)^m,
\end{equation}
where $m = 2$ ($m$ must be even for bosons and odd for fermions). 
In this expression we have 
only written the {\it polynomial} part of the 
wave function. We have omitted (and will omit henceforth) 
a factor that includes the normalization constant and a product of 
single-particle orbitals involving the gaussians that
depend on $|z_i|^2$ and the orbital that describes the state along the rotation axis. 
By demanding all particles to be in the LLL,
the above function can  only  
depend on $z_j$, and not on its complex conjugate, $\bar{z}_j = x_j - i y_j$. To understand
this physically, notice that in the LLL the kinetic energy 
is minimized while the angular momentum is
maximized. Therefore, the  
angular momentum of every
particle  along the rotation axis  
must be positive or zero. This is
true as long as  the polynomial part of 
the wave function depends  {\it only} 
on    $z^{l}_j \propto e^{il\theta_j}$, 
with $l\geq 0$.
In other words, the polynomial part
of the wave function  
must be  an {\it analytic} 
function of $z_j$, for $j=1,\ldots,N$.
Therefore, allowing $\sum_{j=1}^{N} H_j$ to act on 
$\Phi_m$ yields $N$ times the energy of a 
particle in the lowest Landau 
and axial levels,  
$\hbar (\omega_{\perp} + \omega_{||}/2)$, 
plus the  confinement energy, $\upsilon L_{\rm o}$,
where $L_{\rm o} = 
\hbar m N(N-1)/2 = \hbar N (N-1)$
is the  total angular momentum of the Laughlin state
 (to see this, consider a rigid rotation of the system where $z_i \to z_i e^{i\theta}$,
the angular momentum can be read off  the  
phase factor of the many-body wave function). Furthermore, 
one can easily  see  that the Jastrow structure of ~(\ref{laughlin}) produces a state 
with  zero interaction energy for $V_{ij} = g_{2d} \, \delta^{(2)}({\bf r}_i - {\bf r}_j)$,
which in this case represents the relevant interaction for ultra-cold 
atoms confined to two dimensions~\cite{Petrov00}.

	Let us now consider the excitations about the 
Laughlin state. Bulk excitations ~\cite{Laughlin83} come in two types:
quasi-holes  and quasi-particles. 
They correspond to a deficit (quasi-holes) or an excess (quasi-particles) 
of $\frac{1}{m} = \frac{1}{2}$ boson in the bulk
of the droplet. Another interesting
property, which will be useful in the following section, is that creating a quasi-hole increases
the {\it average} angular momentum of the system, whereas creating a quasi-particle
decreases it. For example, by creating a quasi-hole at the center of the 
a droplet: $\prod_{i=1}^{N} z_i \Phi_m$,  so that the resulting state
is an eigenstate of $L$, the angular momentum increases by 
$\hbar N$, and the energy (in the rotating frame) by $\hbar \upsilon N$.
However, if we consider wave functions 
of the form~\cite{Haldane91,Stone92} $s_l \Phi_m$,
where the factor is a symmetric polynomial, $s_l = \sum_{i=1}^{N} z^l_i$, with 
$0 \leq l \ll N$, the angular momentum increases only by $\hbar l$ and
the energy by $\hbar \upsilon l$, much less than the quasi-hole energy. 
Thus we see that the states created by
multiplying the Laughlin state by (products of) $s_l$ are indeed the low-lying
excitations of the system. These states 
describe deformations of the droplet boundary, which in the ground state has a circular shape. This 
is true  for all $l>1$ except for $l = 1$, as  
$s_{1} = \sum_{i=1}^N z_i$ 
represents a small  translation of the droplet center of mass,
which leaves the shape of the boundary unchanged. 

 On the other hand, quasi-particles 
are not the only kind of excitations that decrease the angular momentum
of the system.  Consider  the states created by the action
of the operators $d_l = \sum_{j=1}^{N} \bar{\pi}^l_j$ ($l = 1, 2, \ldots$) 
on the Laughlin state, where $\bar{\pi}_j$
has been defined in Eq.~(\ref{pibar}). According to the discussion that 
we made in previous section, it is easy to understand 
what these operators do: They move up one particle by $l$ Landau levels. 
At the same time, one can show that they {\it decrease} the total angular
momenta by $l$ quanta.  The one-body part of the Hamiltonian  thus
yields an excitation energy (in the rotating frame) 
equal to $\hbar(\omega_{\perp} + \Omega)l$, but strictly speaking the states are not exact
eigenstates of the full many-body Hamiltonian since they do not diagonalize 
the interaction potential. However, their interaction energy for 
$l \ll N$ is much smaller than their one-body energy since 
$gn \ll \hbar \Omega$, and therefore
they can be considered as good approximations to the exact eigenstates. Furthermore,
one can show that for $l=1$, $d_1 \sim \sum_{j=1}^{N} \bar{z}_j = \bar{s}_1$, which 
has zero interaction energy. Thus $d_1$ and $s_1$  are the two independent
modes that describe the center-of-mass motion in the plane perpendicular to the
rotation axis. They are sometimes known as ``Kohn modes'', and their properties 
are discussed in detail in appendix~\ref{appa}. Returning to the general properties of 
the states created by $d_l$, and applying the relation $E_{ROT} = E_{LAB} - \Omega L$
(see appendix~\ref{appa}), we find that  the energy of these states in the laboratory 
frame is $\approx \hbar \omega_{\perp}l$, which means that they are practically degenerate (in the laboratory frame) 
with the surface modes described by $s_l$. This forces us to clarify what we meant above by
``low-lying'' excitations of the system. First of all,  it is necessary to recall  that
the system is at thermal equilibrium in the rotating frame and {\it not} in the laboratory
frame. The characterization of excited states as low-lying or otherwise high-energy states
must be done in this frame, where the states
$d_l \Phi_m$  have much higher energy than the states $s_l \Phi_m$. Therefore, 
at the temperatures $T \ll \hbar \Omega$ of interest here, the statistical probability of finding the system
in the latter states is much higher than  in the former, and therefore the low-temperature
properties are dominated by the states generated by $s_l$. Nevertheless, 
this does not mean that if the system
is probed by an external field of frequency $\omega \approx \omega_{\perp} l$ ($l=1,2,\ldots$) 
which excites modes of either chirality with equal probability,
these excitations are not important to understand the response of the vortex liquid. 
As the frequency of the external field is determined by an apparatus located in the laboratory
frame,  modes of both chiralities will be excited.  This is a distinct feature of 
vortex liquids  compared to the physics exhibited by electronic
quantum Hall systems.  However, as  it will be shown below,  the response to an external probe 
is weighted by a thermal factor  $\left[1 - \exp{(-\hbar \omega_{ROT}(l)/T})\right]^{-1}$,
and at   temperatures such that $\hbar \Omega \gg T \gg \hbar\upsilon$, which should
be more easily accessible, it yields a much higher intensity for the surface modes $s_l$ 
than for the $d_l$ modes. Actually, the considerations just made regarding the $d_l$ modes 
also apply to more general (vortex-liquid) states than the Laughlin state. With these remarks, we close the 
discussion of the excitations generated by the operators $d_l$, and
we shall not discuss them any further in this work.
Instead, we shall focus on the surface modes generated by $s_l$, which exist only when the total
angular momentum is increased by a small number of quanta $l \ge 0$, and for which all
particles remain in the LLL.

 In order to compute the low-temperature properties of a droplet of vortex liquid, 
it is convenient to develop an effective  
field theory that describes the low-energy part of the spectrum.
The states generated by the action of  $s_l$ on $\Phi_m$ have 
a number of  important properties which must be captured 
by the effective theory.  First of all,  they are all {\it chiral} since
$l \geq 0$ and cannot be negative for otherwise the resulting state
would not be in the LLL. Second, the linear dependence on $l$ 
of the  energy implies that the spectral degeneracy is given by $p(l)$, the 
number of partitions of $l$.  This is the number of 
distinct ways $l$ can be written as a sum of  non-negative integers. To
understand this, recall that the angular momentum, and hence
the energy, are proportional to the degree of the homogeneity of 
the polynomial part of the wave function. Therefore, given that $s_l \Phi_m$ 
has an excitation energy equal to $\hbar \upsilon l$, 
so do the states  $s_{l-1} s_1 \Phi_m$ ($l = (l-1) + 1$),  or 
$s_{l-2} (s_1)^2 \Phi_m$ ($l = (l-2) + 1 + 1$), etc.  In the following,
we will show that these features  are  in fact captured by 
quantizing a two-dimensional liquid
drop model~\cite{Wen92}.
\begin{figure}[ht]
\begin{center}
\includegraphics[width=12cm,height=4cm]{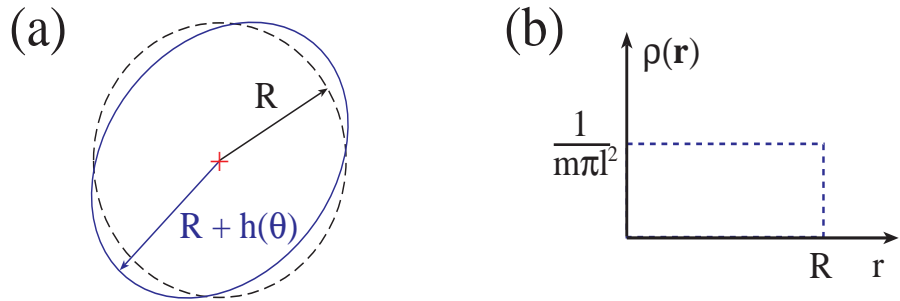}
\end{center}
\caption{(a) Deformation of the droplet boundary and (b) semiclassical 
approximation to the ground state density profile.}
\label{fig2}
\end{figure}

  Following Wen~\cite{Wen92}, we develop an effective low-energy description 
for a finite droplet of $\nu = \frac{1}{m}$-Laughlin liquid. Henceforth, we shall
consider only the Hamiltonian in the rotating frame. However, 
when computing physical properties we shall transform to the laboratory frame (see
Appendix~\ref{appa} for details of how this transformation must be performed).
In the ground state, the shape of the droplet is circular, and the density will be taken to
be uniform and equal to $\rho_{\rm o} = N/A$, $A = \pi R^2$ being the area of the
droplet (see Fig.~\ref{fig1}). We find $A$ by noticing that if $\Phi_m$ is expanded in powers
of, say, $z_1$, the highest power is $l_{\rm max} = m(N-1)$, which corresponds
to the angular momentum of the highest occupied single-particle
orbital in the LLL. The extent  of this orbital gives the radius of the droplet,  
$R = l^{1/2}_{\rm max}\: \ell \approx \left(mN\right)^{1/2} \: \ell$;
hence $\rho_{\rm o} = 1/(m\pi \ell^2)$ for $N \gg 1$. Next, imagine a deformation 
of the droplet such that the boundary is shifted to $R + h(\theta)$ for $0 \leq \theta < 2\pi$.
Clearly, since this is a low-energy description the deformation must be small:
$h(\theta) \ll R$. In addition, our hydrodynamic description necessarily breaks down
at the scale of the oscillator length $\ell$. This implies that 
$h(\theta) \gg \ell$ or, in terms of angular momentum of the excitations, 
the effective theory will be valid for modes with angular 
momentum $\lesssim \hbar \sqrt{mN}$.

  In what follows, we  work with the particle density per unit 
angle (sometimes we shall call it {\it current} because it is 
proportional to it), 
\begin{equation}
j(\theta) = \rho_{\rm o} R h(\theta).
\end{equation}
Let us write down the energy cost (i.e. the Hamiltonian) 
for a deformation of the droplet, being 
$v({\bf r}) = M(\omega_{\perp}^2 - \Omega^2)\: {\bf r}^2/2$, the confining 
potential in the rotating reference frame (cf.~ Eq.~(\ref{eq2})), and
$\mu \equiv v(|{\bf r}| = R)$, the chemical potential. The appropriate 
expressions  read:
\begin{equation}\label{hameff1}
H = \int d^2 {\bf r}\, \left(v({\bf r}) -  \mu \right) 
\left[  \rho({\bf r}) - \rho_{\rm o}(r) \right]
=   \frac{F R\rho_{\rm o}}{2} 
\int^{2\pi}_{0} d\theta \, h^2(\theta) =
 \frac{F}{2\rho_{\rm o}R} \int^{2\pi}_{0} d\theta \, j^2(\theta),
\end{equation}
where $\rho_{\rm o}(r) = \rho_{\rm o} \theta(R-r)$ and 
$F = v'(R)$. We will  allow
for the possibility that the deformation involves the addition or removal
of a small number $Q = \int^{2\pi}_{0} d\theta \, j(\theta) \ll N$ of particles 
at the boundary. Canonical quantization of~(\ref{hameff1}) can 
be achieved with the help of the continuity equation, 
\begin{equation}
\partial_t  j(l,t)  =  i \upsilon  l\:  j(l,t),
\end{equation}
which has been expressed in terms of 
the Fourier modes of $j(\theta) = \sum_{l} j(l) \exp\left(-i l \theta \right)/2\pi$.
Notice that, contrary to the case of single-particle orbitals, in this case $l$ 
measures, not the absolute angular momentum of the excitation, but
the increase in the angular momentum over the  ground state.
The quantized effective Hamiltonian  takes the form
\begin{equation}\label{hameff2}
H = \hbar \upsilon m \sum_{l>0} \: j(l) j(-l) + \frac{\hbar \upsilon}{2} m Q^2.
\end{equation}
The canonical commutation relations  lead to the following
{\it current  algebra}:  $\left[ j(l), j(l') \right] = \frac{l'}{m} \delta_{l+l',0}$, 
which in the  mathematical Physics  literature is 
known as U$(1)$ Kac-Moody (KM) algebra.  Nevertheless, 
both the Hamiltonian and the current algebra can be written 
in a more familiar way if, for $l > 0$, 
we introduce  $b(l) \equiv -i \sqrt{\frac{m}{l}}\, 
j(-l)$ and $b^{\dagger}(l) \equiv i \sqrt{\frac{m}{l}}\, j(l)$;
hence $\left[b(l), b^{\dagger}(l') \right] = \delta_{l,l'}$, i.e. the familiar
``phonon'' algebra. In terms of these operators 
\begin{equation}\label{hameff}
H = \sum_{l > 0} \hbar \upsilon l \, b^{\dagger}(l) b(l) +   \frac{\hbar \upsilon}{2} m Q^2,
\end{equation}
i.e.  the Hamiltonian of a system of {\it chiral} phonons with energy 
dispersion $\hbar \upsilon l$ ($\hbar \omega_{\perp} l$ in the laboratory frame). 
Therefore, the spectral degeneracies are 
given by $p(l)$.   The chirality of surface modes 
stems from  the broken time-inversion
symmetry existing in this situation: The system rotates in one direction, which is reversed
under time inversion. Furthermore, it is also important 
to point out that the state annihilated  by the $b(l)$ operators
must be the Laughlin state. This is related to  the existence, {\it in the rotating frame},
of a finite energy gap which must be overcome in order to decrease the angular
momentum  below that of the Laughlin state, and should be interpreted as due to 
the incompressibility  of the Laughlin state (i.e. the droplet cannot further shrink)~\cite{Lee91}.
Thus  we conclude that the effective theory correctly describes all
the features of the low-energy spectrum obtained using wave functions. 

	The first  property that we want to compute using the effective
theory is the response to a weak time-dependent  deformation of the 
the trapping potential. 
This couples to the particle density at the boundary,
and therefore to $j(\theta)$ (see appendix~\ref{appa} for more details).  The measurable 
response will be  given in terms of the dynamic structure factor, 
\begin{equation}
S_{ROT}(l,\omega) = \int_{-\infty}^{+\infty} dt \: e^{i\omega t} \,  \int _{0}^{2\pi} d\theta \: e^{-il\theta} 
\langle j(\theta,t) j(0,0) \rangle,
\end{equation}
where the brackets stand for thermal average over the canonical ensemble in the rotating
frame, which is where the system is in thermal equilibrium. The
above function can be written in terms of the response to a external potential,
by virtue of the fluctuation-dissipation theorem~\cite{Forster}.
At a temperature $T$, the relationship is given by
\begin{equation}
S_{ROT}(l,\omega) =  \frac{2 \hbar 
\: {\rm Im} \: \chi(l,\omega)}{\left(e^{-\hbar \omega/T}-1\right)},  
\end{equation}
where $\chi(l,\omega)$ is the Fourier transform of 
\begin{equation}
\chi(\theta,t) =  -\frac{i}{\hbar}\vartheta(t)\:  \langle  \left[ j(\theta,t), j(0,0) \right] \rangle,
\end{equation}
where $\vartheta(t)$ is the step function. 
The previous correlation function can be readily computed 
using current algebra described above and  the continuity equation, which 
implies that $j(\theta,t) = j(\theta-\upsilon t)$.
Thus, applying the results of appendix~\ref{appa}, we arrive at the following result
for the dynamic structure factor in {\it laboratory} frame:
\begin{equation}\label{dsf}
S_{LAB}(l,\omega) = S_{ROT}(l,\omega-l\Omega) = 
\frac{l}{m} \: \frac{\delta(\omega -  \omega_{\perp} l)}
{1-e^{-\hbar (\omega_{\perp} - \Omega) l/T}}.
\end{equation}
Since there is a single branch of phonons,  the fact that $S_{LAB}(l,\omega)$  is peaked 
at $\omega =  \omega_{\perp} l$ should not be surprising (in real systems,
a finite broadening of the peak is expected). However, 
it is also important to notice that  the spectral weight of the peak associated with the 
surface mode (cf.  Eq.~(\ref{dsf}))  is proportional to the 
filling fraction $\nu = 1/m$ of the Laughlin state. 

A way to understand the proportionality   to
the filling fraction of the spectral weight  is to show that, as far as
the response to an external potential is concerned, a
droplet  of Laughlin liquid behaves 
as a rapidly rotating cloud of non-interacting particles (i.e. an ideal gas) with  {\it peculiar exclusion
statistics}~\cite{Haldane91}. To understand what this means, let us assume that when
the ideal gas is in thermal equilibrium it is described by a 
density matrix $\rho^{0}$ such that $\rho^{0} | k \rangle = N(k) | k\rangle$.
The function $N(k)$ is the occupancy 
of a single-particle orbital $| k \rangle$ with
angular momentum equal to $\hbar k$. 
To compute the {\it linear} response of this ideal gas to the 
external perturbation $\delta v_{ext}(\theta,t)$ localized near the boundary, 
let us consider the linearized equation of 
motion for the perturbed density matrix $\rho^{0} + \delta\rho(t)$,
\begin{equation}
i\hbar   \langle  k + l | \partial_{t}\delta \rho(t) | k \rangle = \langle  k + l | 
\left[ H_{\rm o}, \delta \rho(t) \right] + \left[ \delta v_{ext}(\theta,t), 
\rho^{0}  \right] | k \rangle. 
\end{equation}
where $H_{\rm o} | k \rangle = \varepsilon(k) | k \rangle$, $\varepsilon(k) = 
\hbar \upsilon k$ being the single-particle dispersion.  We thus find ($\eta \to 0^{+}$):
\begin{equation}
\chi(l,\omega) = \lim_{\delta v_{ext} \to 0} \, \frac{\delta 
\langle j(l,\omega) \rangle}{\delta v_{ext}(l,\omega)} =
\frac{1}{2\pi \hbar}\:  \sum_{k} \frac{N(k) - N(k+l)}
{\omega - \left( \varepsilon(k+l) - \varepsilon(k) \right)/\hbar + i\eta}.
\end{equation}
Since the particles have been assumed to be non-interacting
we can use the results  of Sect.~\ref{disgr}, which imply that 
the single-particle dispersion is {\it linear} with
the angular momentum. Thus the previous expression reduces to
\begin{equation}
\chi(l,\omega) = \frac{1}{2\pi \hbar} \frac{1}{\omega - \upsilon l + i\eta} \:
\sum_{k=0}^{+\infty} \left[ N(k) - N(k+l) \right] = \frac{g^{-1}}{2\pi\hbar} \:
 \frac{1}{\omega - \upsilon l + i\eta}.
\end{equation}
It is important to stress that the above expressions make
sense as long as $l \ll \sqrt{mN} $, otherwise the effect of the external potential is not 
reduced to the neighborhood of the boundary and radial part of the single-particle 
orbitals must be taken into account. In the previous expression
we have denoted as $g^{-1}$ the mean occupancy at the bottom of 
the single-particle band (cf. Fig.~\ref{fig3}), 
which at the low temperatures where the gas is quantum degenerate 
must be a constant. By  comparing 
this expression for $\chi(l,\omega)$ with the one obtained from the effective
theory, we arrive at the identification $g = m = 2$, which 
implies that at the bottom of the band the mean occupancy is $\frac{1}{2}$.
This may seem surprising, as it means that near the bottom of the band
there is one particle per two states, a situation that does not correspond to neither
fermions ($g = 1$) nor bosons ($g = 0$)~\cite{Haldane91}. However, we can argue that 
this result makes  indeed sense. Let us first emphasize that we have
assumed that the droplet is  an ideal gas, which seems to be at
odds with the fact that the actual particles (i.e. the bosons) are  interacting. However, we must remember
that under the combined action of  interactions and  rapid rotation, the bosons effectively
become {\it hard-core} so that their many-body wave function has zero interaction
energy. This pushes the atoms to orbits where their relative 
angular momentum is equal to $2\hbar$. As a result,
there is, on average, one boson per every two angular momentum states: We
have $N$ particles in the states from $l=0$ to $l_{max} = 2(N-1)$, i.e. 
in $2N-1$ states, which for $N \gg 1$ yields two states per particle. Therefore,
the rapid rotation plus  interactions become a ``statistical'' interaction, which makes
the bosons behave as if they are non-interacting 
``super-fermions'' obeying Haldane's {\it exclusion}
statistics~\cite{Haldane91,Wu94,Viefers00} with  $g = 2$. 
Going back to  our discussion of the experimental characterization of
the Laughlin state, we have thus shown that the appearance 
of the filling fraction in the  spectral weight of $S_{LAB}(l,\omega)$ 
can be interpreted as an sign of the peculiar   
statistical aspects of the Laughlin liquid. Therefore, 
if one measured  the spectral
weight of $S_{LAB}(l,\omega)$ and found it to be proportional to
$\frac{1}{m}$, this would provide a fairly direct 
evidence for the fact that exclusion statistics is at play
in the Laughlin state~\cite{Viefers00,Elburg98}.  

	Another  function of experimental  interest is
the one-body density matrix. To compute it, one
needs to find  a representation for the bosonic field operator 
at the boundary. To this purpose, it 
will be convenient to introduce the phonon field $\phi(\theta)$, related to
the density $j(\theta)$ by  $\partial_{\theta} \phi(\theta) = 2\pi j(\theta)$. 
Expanding it in normal modes: 
\begin{equation}\label{modes}
\phi(\theta) = \frac{\phi_{\rm o}}{m} + Q\theta + \frac{1}{\sqrt{m}}\sum_{l > 0} 
\frac{1}{\sqrt{l}}    \left[ e^{il\theta} b(l)  +  e^{-il\theta} b^{\dagger}(l) \right],
\end{equation}
with $[Q,\phi_{\rm o}] = i$. In order to construct an operator with the same
properties as the boson field operator, we first notice that if a boson
is added at $\theta_{\rm o}$, the density $j(\theta)$ 
becomes $j(\theta) + \delta(\theta - \theta_{\rm o})$.
By direct calculation using~(\ref{modes}), one can check that  
$[j(\theta), \phi(\theta')] = \frac{i}{m} \delta(\theta - \theta')$, which means
that $j(\theta) $ and $m \phi(\theta)$ are canonically
conjugate to each other.  Thus the operator that we
seek must be proportional to $e^{-im\phi(\theta)}$, since
$e^{im\phi(\theta_{\rm o})} j(\theta) e^{-im\phi(\theta_{\rm o})} = 
j(\theta) +  \delta(\theta - \theta_{\rm o})$. In this 
construction, the boson appears as 
a {\it soliton} or {\it kink} in the field $\phi(\theta)$ with topological charge $Q = + 1$. 
Fractionally charged   excitations can be also constructed, though they
are not physical boundary excitations for the droplet geometry. They
are created  by the operator $e^{-i\phi(\theta_{\rm o})}$, which shifts 
$j(\theta) \to j(\theta) + \frac{1}{m} \delta(\theta-\theta_{\rm o})$
and therefore describes the creation of a quasi-particle at $z_{\rm o} 
\approx R e^{i\theta_{\rm o}}$.  One important consistency
test, before we proceed any further, is to show that the putative boson operator,
and its hermitian conjugate, are commuting at different points. 
For $\theta \neq \theta'$, we have that
\begin{eqnarray}
e^{im\phi(\theta)} e^{\pm i m \phi(\theta')} &=& e^{\mp m^2 \left[ \phi(\theta), 
\phi(\theta') \right]} e^{\pm im\phi(\theta')} e^{ i m \phi(\theta)}\nonumber \\
&=& (-1)^{m} e^{\pm im\phi(\theta')} e^{ i m \phi(\theta)},
\end{eqnarray}
and $(-1)^m = +1$ since $m = 2$. 
Indeed, by repeating this calculation with $m$ replaced by $\nu^{-1}$, one can 
see that  commuting fields are obtained {\it only} when $\nu^{-1} = m$ is an 
even integer (if $m$ is odd, the fields anti-commute, and this is the situation
usually encountered in the FQHE, where the constituents are electrons). 
Wen has argued~\cite{Wen92} that  for $\nu \neq 1/m$, 
there must be more than one phonon branch in the spectrum,
otherwise no field operator can be constructed
and the effective theory is not self-consistent. Below, we shall
give explicit examples of how the additional branches  appear.
To close this part of the discussion, it is worth mentioning 
that the {\it exchange} statistics of the excitations
created by  $e^{-i\phi(\theta)}$ is 
fractional, since $e^{-i\phi(\theta)} \: e^{-i \phi(\theta')} = 
e^{-i\pi {\rm sgn}(\theta-\theta')/m} \, e^{-i\phi(\theta')} \: e^{-i \phi(\theta)}$.
That is,  after exchanging two boundary quasi-particles  we
pick up a phase factor equal to $\pm \pi/m = \pm \pi/2$. These excitations
are $\frac{1}{2}$-{\it anyons}.

The complete form 
of the (boundary) field operator in  terms of the phonon field is
\begin{equation}\label{fieldop}
\Psi^{\dag}(z=Re^{i\theta})\equiv
\Psi^{\dagger}(\theta) = A\: e^{-im \left[ l_{\rm o}\theta + \phi(\theta) \right]}, 
\end{equation}
where $A$ is a constant that depends
on the way the modes at high  $l\gtrsim \sqrt{mN}$ are excluded from the above sums.
The value  $l_{\rm o} = (N-\frac{1}{2})$ is obtained by comparing with the spectral
representation of the one-body density matrix, Eq.~(\ref{dm}). 
Calculation of the one-body density-matrix at the boundary is now possible 
using~(\ref{modes}) and~(\ref{fieldop}):
\begin{equation}\label{dm}
G(\theta) = \langle \Psi^{\dagger}(\theta) \Psi(0) \rangle = 
e^{-iml_{\rm o}(\theta-\theta')}\: \langle e^{-im\phi(\theta)} e^{im\phi(0)} \rangle.
\end{equation}
\begin{figure}[ht]
\centerline{\resizebox{9cm}{!}{\includegraphics{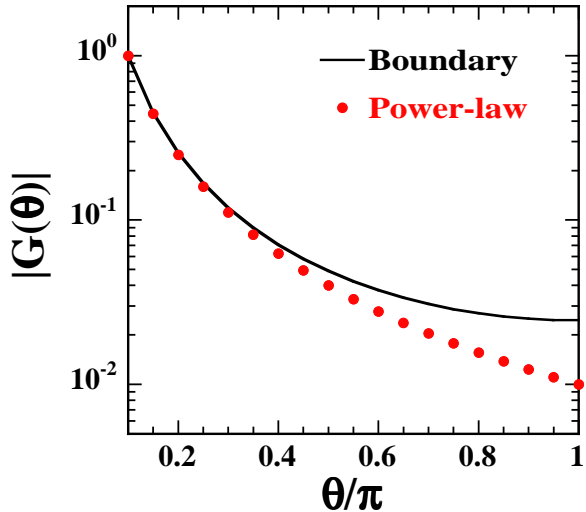}}}
\caption{Behavior of the density matrix at the boundary compared with
the power-law $1/\theta^2$. The functions
have been normalized to $1$ at $\theta = 0.1 \pi$. The power-law decay
is accurately followed for small angles, whereas for large angles
$|G(\theta)|$ decays more slowly since it is periodic on a finite-length boundary. 
In this respect, notice that $|G(\theta)|$  has reflection 
symmetry about $\theta = \pi$.}
\label{fig5}
\end{figure}
This function has two interesting limits ($\theta \gg \ell/R$ in both cases). 
When the temperature is higher than the phonon 
level spacing, i.e. if $T \gg \hbar \upsilon$, then 
\begin{equation}
G(\theta) = {\rm const.} \times \left( \frac{\pi T}{\hbar \upsilon}\right)^m 
 e^{-iml_{\rm o}\theta} \left[  \sinh \left(\frac{\pi T \theta}{\hbar \upsilon} \right) 
\right]^{-m},
\end{equation}
and therefore correlations decay exponentially at large $\theta$, but have
power law form at small $\theta$.  In the more demanding temperature 
regime where $T \ll \hbar \upsilon$, the density matrix reads:
\begin{equation}\label{dmt0}
 G(\theta) = {\rm const} \times e^{-iml_{\rm o}\theta}  \left[ \sin \left(\frac{\theta}{2}
\right) \right]^{-m},
\end{equation}
and  exhibits an ``almost power-law'' behavior with $\theta$ (see Fig.~\ref{fig5}), which is cut off
by the finite length of the boundary.  The last form, Eq.~(\ref{dmt0}),  
has been  numerically shown to be accurate for  a 
system of $N = 36$ bosons by Lee and Wen~\cite{Lee91} (see also 
Ref.~\onlinecite{Mandal02} for $m = 3$ calculations). It is interesting to
compare Eq.~(\ref{dmt0}) with  the one-body density matrix 
away from the boundary~\cite{MacDonald88},
\begin{equation}\label{dmbulk}
G({\bf r}, {\bf r'}) = \frac{1}{m\pi\ell^2} e^{-|z-z'|^2/2\ell^2} \, e^{(z^* z'-z'^* z)/2\ell^2}.
\end{equation}
Apart from the last term, which is a phase factor, 
it can be seen that this function
decays at large distances as a gaussian. This is in contrast to the
almost-power-law decay that the same function 
exhibits near the boundary, where  $|{\bf r}| \approx |{\bf r'}| \approx R$. The
differences in behavior are due to the {\it quantum critical} fluctuations of
the surface modes.
\begin{figure}[t]
\centerline{\resizebox{9cm}{!}{ \includegraphics{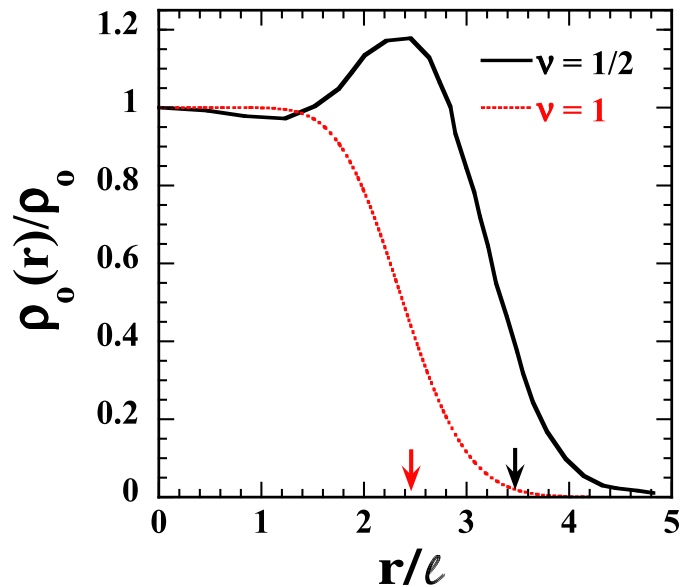} } }
\caption{Density profile of the $\nu = 1$ and $\nu = \frac{1}{2}$ Laughlin 
states for $N=6$
particles. The latter corresponds to boson and the former to fermions. 
The density is normalized to the bulk
values $\rho_{\rm o} = 1/\pi \ell^2$ for the $\nu = 1$ 
and $\rho_{\rm o} = 1/2\pi \ell^2$ for $\nu = \frac{1}{2}$. 
Notice that the $\nu = \frac{1}{2}$ state is more extended than
the $\nu = 1$. The arrows indicate the position of the semiclassical radius 
$R = \sqrt{m N}\: \ell$. Data for the $\nu = \frac{1}{2}$ state has been taken from
Ref.~\onlinecite{Paredes02}.}
\label{fig1}
\end{figure}

  Nevertheless, it seems difficult that the behavior 
of density matrix near the boundary,  Eq.~(\ref{dmt0}), can be experimentally measured.
This is because  for a small droplet any measurement of the one-body correlation function
would be dominated by the bulk signal 
coming from~(\ref{dmbulk}). However, it turns out that the result~(\ref{dmt0})
does have some effect on the density profile. To see this, we first notice 
that Eq.~(\ref{dmt0}) can be written using the binomial expansion as
\begin{equation}
G(\theta) = {\rm const.} \times \sum_{p=0}^{+\infty} \frac{(p + m - 1)!}{p! (m-1)!}\,  
e^{i (p - l_{max})\theta},
\end{equation}
from which the occupancy of the levels near $l_{max}$ can be obtained:
\begin{equation}
\langle n(l) \rangle \propto \int_{0}^{2\pi} d\theta \, G(\theta) e^{il\theta}
\end{equation}
with $l = 0, 1, \ldots$ Thus we find~\cite{Wen92} that, for $l \leq l_{max}$,
\begin{equation}\label{occ}
\langle n(l) \rangle = {\rm const.} 
\times \frac{( l_{max}  - l + m - 1)!}{(l_{max} -l)! (m-1)!}
\end{equation}
and vanishes for $l > l_{max}$, as expected. Therefore, for $m = 2$, 
the occupancy near the boundary behaves as $(l_{max} + 1 - l)$. 
But given the limitations of the effective theory, this result
will valid for $(l_{max}-l) < \sqrt{mN}$. Mitra and MacDonald~\cite{Mitra93}
computed  $\langle n(l) \rangle$ numerically for small Laughlin droplets with $m= 3, 5, 7$
and found good agreement with the general expression, Eq.~(\ref{occ}), within
its validity range. Moreover, they also found that $\langle n(l) \rangle$ exhibits 
a prominent peak followed by smaller oscillations 
as $l$ decreases from  $l_{max}$ to $0$, i.e.
as one moves from the boundary to the center of the droplet.
For large enough droplets, the oscillations eventually damp 
out as $\langle n(l) \rangle$  approaches, rather slowly, the average 
value of $\frac{1}{m}$. This behavior can be understood
on the basis of two facts: i) The existence of a cut-off
at zero temperature,  for $l = l_{max}$, above which $\langle n(l) \rangle$ vanishes  
according to Eq.~(\ref{occ}) (e.g. as $(l_{max} + 1 - l)$
for $m=2$), and ii) the fact that the average occupancy 
equals $\frac{1}{m}$ ($= \frac{1}{2}$ in the present case).
These two facts imply that as $l$ approaches $l_{max}$ from below, the occupancy
must necessarily {\it decrease} below the average $\frac{1}{2}$.
However, to  maintain the average occupancy, the particles 
removed from the neighborhood of $l_{max}$ must be 
placed in orbitals with lower angular momentum. Since for
relatively large droplets, $\langle n(l) \rangle \approx \frac{1}{m}$ for $l \ll l _{max}$, 
it seems reasonable to expect that $\langle n(l) \rangle$ exceeds the average
by displaying a maximum before it decays 
to zero as required by Eq.~(\ref{occ}).
As the orbitals with $l$ quanta of angular momentum 
are located  around $r_l \sim \ell \sqrt{l}$, 
the  peak in $\langle n(l) \rangle$  translates into a peak in the  density 
near the boundary. This  can be observed in Fig.~\ref{fig1},
where the density profile of a $\nu = \frac{1}{2}$ Laughlin
droplet with  $N = 6$ particles has been plotted.
The density is related to $\langle n(l) \rangle$ 
by the following expression,
\begin{equation}\label{density}
\rho_{\rm o }({\bf r}) = \sum_{l=0}^{l_{max}} 
\langle n(l)  \rangle\,  |\varphi_l ({\bf r})|^2.
\end{equation}
This result is a consequence of the Laughlin state being an eigenstate
of the total angular momentum. When summing over $l$,
the occupancy at angular momentum $\hbar l$ 
is averaged over the neighboring orbitals, and in the present
case  any oscillations displayed by $\langle n(l) \rangle$ are washed out.
The density near the center of the trap reaches the 
constant value of $1/(2\pi \ell^2)$, which corresponds to
the average occupancy $\frac{1}{2}$. Nonetheless, a 
peak in the density near the boundary appears and should  visible in
an experiment where the density profile is measured.
In this respect, it was shown  in Ref.~\cite{Ho02} using 
Eq.~(\ref{density}) that by turning the trapping potential off
and allowing the atoms to expand freely, the density profile
of a vortex liquid expands self-similarly. This fact can be used
to perform very accurate measurements of the density profile,
which should reveal a characteristic peak near the boundary 
for the Laughlin state (and possibly for other vortex liquids), and 
which is not expected for Bose-condensed systems.

  For comparison purposes, in Fig.~\ref{fig1} we have also plotted the density profile
for the fermionic $\nu = 1$ Laughlin state. This is a state where
particles are non-interacting, and  the occupancy is known  exactly: $\langle n(l) \rangle 
= 1$ for $l \leq l_{max} = N-1$,
and zero for $l > l_{max}$. This is again in agreement with~(\ref{occ}),
which predicts  $n(l) = {\rm const.}$, for $l \leq l_{max}$ (this
constant cannot be fixed by the effective theory).  In this case
$\langle n(l) \rangle$ vanishes abruptly for $l = l_{max} + 1$. 
Therefore, we do not expect a
peak  in the density near the boundary.
We conclude that the peak is a consequence 
of the strong correlations built into
the Laughlin state.  It  is indeed a generic
property of finite droplets of the Laughlin liquid, and in a more general
framework is   a particular
example of the generalized Luttinger's theorem~\cite{Luttinger60}
discussed by Haldane in Ref.~\onlinecite{Haldane92}.

 We end this section by emphasizing the fundamental 
difference between the boson
occupancy, $\langle n(l) \rangle$, and the occupancy for 
the ideal ``super-fermion'' gas, $N(l)$, introduced in our discussion of the
density response of the Laughlin droplet. The latter is a convenient tool,
which can be also used to discuss some other properties of the
Laughlin liquid, e.g. the thermodynamics of the 
boundary excitations or the Hall conductance~\cite{Elburg98}. The boson
occupancy $\langle n(l) \rangle$  contains in addition 
correlation effects between the actual particles, 
which lead, for instance, to the almost 
power-law behavior of $G(\theta)$. A mathematically rigorous 
discussion of these issues can be found in Ref.~\onlinecite{Elburg98}.


\section{Other stable states of scalar bosons}\label{other}

\begin{figure}[t]
\centerline{\resizebox{12cm}{!}{\includegraphics{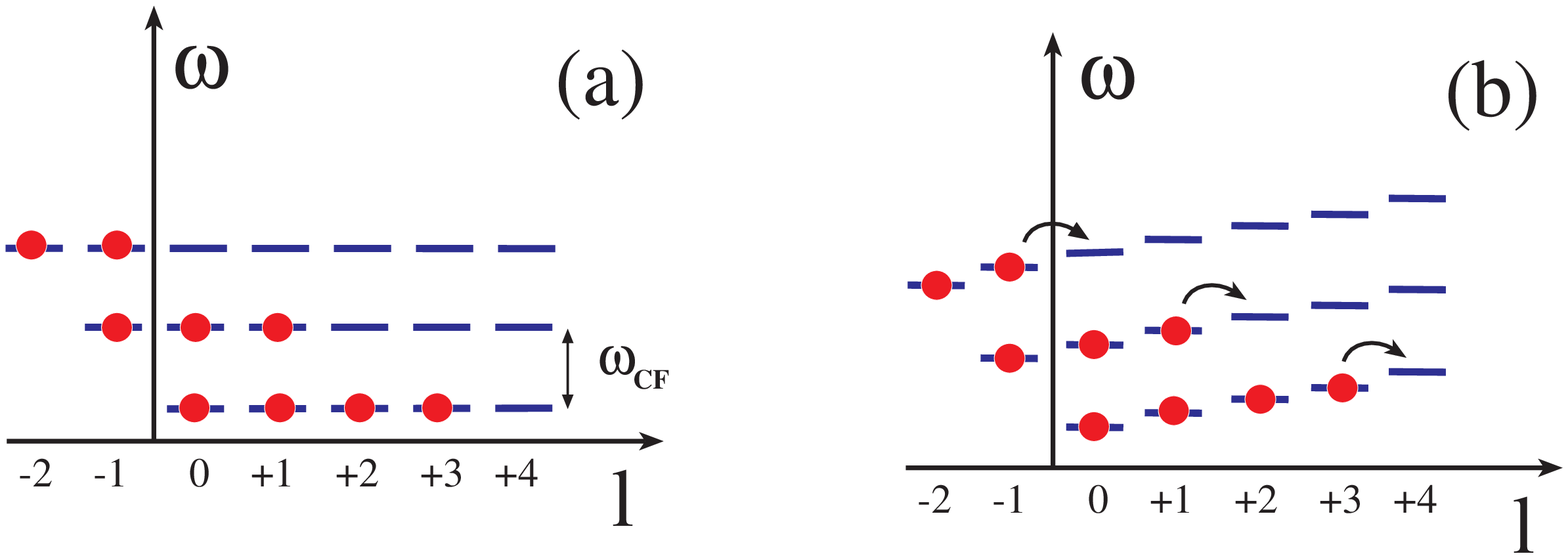}}}
\caption{A  compact state with $N =9$ composite fermions (a and b), which corresponds to the
integer sequence $[4,3,2]$.  In (b) we have represented the effect of the confinement, which is
to lift the degeneracy of the otherwise degenerate levels within every CF Landau level. Surface
excitations can be pictured as low-energy particle-hole excitations (b)  within each Landau level.
For these excitations  the composite fermions can be weakly interacting, even if
the interactions could be neglected in the ground state.}
\label{fig4}
\end{figure}

 The Laughlin state described in the previous section has the property of being
an exact  eigenstate of the Hamiltonian:
\begin{equation}\label{hamtot}
H = \sum_{i=1}^{N} H_j + g_{2d} \sum_{i<j=1}^{N} \delta({\bf r}_i - {\bf r}_j).
\end{equation}
More precisely, (\ref{laughlin}) is the state with  the 
lowest energy for $L = \hbar mN(N-1)/2$. On the other hand, 
the states that we will consider in this section are not exact eigenstates 
of~(\ref{hamtot}). They have been obtained numerically using two different
types of  exact diagonalization set-ups. If one is only interested
in the bulk properties of  vortex liquids,  and taking into account
that exact diagonalization is only feasible with
relatively small particle numbers, the Hamiltonian can be
efficiently   diagonalized on a {\it compact} manifold, like 
a torus~\cite{Cooper01}  or a sphere~\cite{Regnault02}. 
Cooper {\it et al.} used the toroidal geometry to study the
stability of the Abrikosov lattice as the filling fraction 
$\nu = N/N_v$ ($N_v$ being the number of vortices) is varied.
They found that the lattice melts for $\nu \sim 6$ and that
for smaller filling fractions a rotating bosonic system exhibits 
a series of gapped states ({\it homogeneous}  vortex liquids) when
$\nu$ belongs to the sequence   $\frac{1}{2}, 1, \frac{3}{2}, 2,  \ldots,\frac{5}{2}
\ldots$ The gap is a signature of the incompressiblity of the 
vortex liquid. The $\nu = \frac{1}{2}$ state was found to correspond
to the Laughlin state discussed above, while the
remaining states were shown to have good overlap with the Moore-Read state
(for $\nu = 1$, see below) and other wave functions of the class introduced 
by  Read and Rezayi~\cite{Read99}. At the same time, 
the Laughlin and  Moore-Read states had been previously found  
in another type of exact diagonalization 
studies~\cite{Cooper99,Wilkin00} where the bosons are confined by
a harmonic potential. This is a situation that is closer to the one experimentally
relevant, and in which the vortex liquids are {\it inhomogeneous}. 
In this set-up, Cooper and Wilkin~\cite{Cooper99} found a series of 
stable states at magic values of the total angular 
momentum, $L$. The series terminates at the $\nu = \frac{1}{2}$-Laughlin state,
and many of remaining the stable states have good overlap with
wave functions constructed from the following ansatz:
\begin{equation}\label{CF}
\Phi^{(n)}_{\rm B}(z_1,\ldots,z_N) = {\cal P}_{\rm LLL} 
\left[ \Phi_1(z_1,\ldots,z_N)  \Phi^{(n)}({\bf r}_1,\ldots,{\bf r}_N) \right],
\end{equation}
where $\Phi_1 = \Phi^{(1)}$ denotes the Jastrow factor, 
$\prod_{i<j}(z_i-z_j)$, and $\Phi^{(n)}$
is a Slater determinant representing the state
of $N$ fictitious fermions, 
the so-called ``composite fermions'' (CF's),
distributed over $n$ Landau levels. The operator 
${\cal P}_{\rm LLL}$ is needed to project 
the ansatz onto the LLL (for details
of how this projection should be carried
see Ref.~\onlinecite{Cooper01} 
and Ref.~\onlinecite{Viefers00}). Furthermore, Cooper and Wilkin
found that the stable states 
are  ``compact'',  a term coined by Jain and Kawamura~\cite{Jain95}
to refer to those states where $N_i$ CF's occupy 
the lowest angular momentum orbitals  
in the  $i$-th  CF Landau level ($i=0,1,\ldots, n-1$, 
so that $\sum_{i=0}^{n-1} N_i = N$, see Fig.~\ref{fig4}).
Therefore, a compact state is 
characterized by the  set of
integers: $\left[N_{0}, N_1, N_2, \ldots, N_{n-1} \right]$,
and it has an energy equal to $\hbar \omega_{\rm CF}
\sum_{i=0}^{n-1} N_{i}(i + \frac{1}{2})$, 
where $\omega_{\rm CF} = \omega_{CF}(N)$  is the effective 
CF Landau level spacing~\cite{Jain95,Cooper99}. However,
it is necessary to stress that  not  all
compact CF states are stable~\cite{Jain95}.
In the large $N$ limit,  the CF ansatz 
yields {\it homogeneous} 
vortex liquids whose filling fraction falls
into the {\it bosonic} Jain sequence~\cite{Cooper01},
i.e. $\nu = n/(n+1)$  or $\nu = (n+1)/n$, for an integer $n\geq 1$. 
Except for the $\nu = \frac{1}{2}$-Laughlin state, this sequence
does not match the one found by Cooper {\it et al.}
using the toroidal geometry, and this has  led to some confusion. 
The situation has been recently
clarified by Regnault and Jolicoeur~\cite{Regnault02} who
performed exact diagonalizations on another compact geometry, namely a sphere,
and checked for the convergence of the spectral gap with the system size. They found 
the existence of stable homogeneous vortex liquids for $\nu = \frac{1}{2},
\frac{2}{3}, \frac{3}{4}, \frac{4}{3}, \frac{5}{4}$, falling into to
the Jain sequence,  and for $\nu = 1$, which corresponds to the  Moore-Read 
state.  Therefore, below we shall consider the surface modes of 
some of the states in the Jain sequence as well as the Moore-Read state.

	To study the surface waves of the vortex liquids whose wave functions 
can be obtained  from the CF ansatz, Eq.~(\ref{CF}),
we will employ the {\it parton} construction~\cite{Wen91,Wen92}.
Usually,  this method assumes that the number of
particles, $N$,  is macroscopically large so that the inhomogeneity of the system  
can be neglected. For these states, one can safely assume that
there are  $n$ CF Landau levels filled by $N/n$ composite
fermions. As pointed out above, the resulting state is characterized by a filling fraction 
$\nu = n/(n+1)$. However, when dealing with a {\it mesoscopic} droplet,  
this is not necessarily true
and the compact states exhibiting good overlap with the stable states 
usually have {\it unequal} number of CF in different Landau levels. 
Generalizing the parton construction for this situation is not difficult,
the only limitation being imposed by the subsequent use  
of the effective low-energy 
theory introduced in Sect.~\ref{sectlaughlin} to 
describe of the surface waves of a given 
CF Landau level, $i$. This requires that the CF level occupancy $N_i \gg 1$, 
which is not  always fulfilled.  Nevertheless,
it was found~\cite{Cappelli98} that having some of the higher CF Landau 
levels occupied with $N_i \sim 1$  CF's corresponds to quasi-particle excitations
over a state of the Jain hierarchy. Therefore,
these states can be characterized by an effective 
number of  CF levels to which the effective theory
can be applied. In the discussion that follows, we denote  
this number by $p$ ($\leq n$).  In other words, this integer, which yields the
{\it effective} filling fraction $\nu = p/(p+1)$, can be associated with
the number of CF Landau levels with occupancy 
$N_{i} \gg 1$ .

 Keeping in mind the above caveats, we can 
regard  the CF ansatz~(\ref{CF}) as describing
a bound state, the fundamental boson,  made 
up of {\it two} kinds of fermions (called ``partons'')  such that
the many-body wave function (we drop 
${\cal P}_{\rm LLL}$ for notational simplicity)  reads:
\begin{equation}
\Phi^{(n)}_{\rm B}(z_1, \ldots, z_{N}) = 
\Phi_1(z^{(1)}_1, \ldots, z^{(1)}_1) \: 
\Phi^{(n)}(z^{(2)}_1, \ldots, z^{(2)}_N)
 \Big|_{ \left\{z^{(1)}_i = z^{(2)}_i =
  z_i\right\}_{i=1,\ldots,N} }. 
\end{equation}
Thus a boson  is a type-$1$ 
parton in the (fermion) LLL, 
which carries a  ``charge'' $q_1 = p/(p+1)$, bound to a second 
parton (type-$2$), which carries charge $q_2 = 1/(p+1)$ so
that $q_{1} + q_{2} = 1$.  Type-$1$  has one branch 
of boundary excitations, whereas type-$2$ partons contribute with
$p \leq n$ branches corresponding to the $p$ CF  levels with $N_i \gg 1$
(to see how this comes about, repeat the
steps that lead to the effective theory, now setting $m = 1$ for type-$1$
partons, and considering $p$ different branches with $m=1$ for type-$2$
partons).  However, not all of these $p+1$ branches are independent. In fact,
the bound state condition implies that  the density of type-$1$ partons cannot
fluctuate independently of that of type-$2$ partons. Therefore,
we must demand that  all {\it physical} operators do not excite density 
fluctuations where both partons behave independently. Mathematically,
if $O$ is a physical operator, then $\left[ \rho_{\rm c}, O \right] = 0$,
with $\rho_{\rm c} = \rho_{0} - \sum_{i=1}^{p} \rho_{i}$, where
$\rho_{0}$ describes the surface modes of type-$1$ partons, and
$\rho_i$ for $i=1,\ldots,p$ those of type-$2$. These operators  
obey the current algebra:
\begin{equation}
\left[ \rho_{i}(l), \rho_{j}(l') \right] = l' \delta_{l+l',0} \,  \delta_{i,j}.
\end{equation}
As mentioned previously,  
there are $p$ independent branches,
which are obtained by requiring the currents to be physical operators,
in the sense introduced above. Thus it is found that one of the independent currents 
corresponds to the total
current $j_{0} = q_1 \rho_{0} + q_2 \sum_{i=1}^{p} \rho_i$, and the 
remaining $p-1$ are given by $j_{i} = \sum_{j=1}^{p} a_{ij} \rho_{j}$ 
($i = 1, \ldots p-1$), where $a_{ij}$ are $p-1$ orthogonal 
vectors such that $\sum_{j = 1}^{p} a_{ij} = 0$. 

 To illustrate how the parton construction works in practice, 
we first notice that one can readily
recover the results of Sect.~\ref{sectlaughlin}
for the surface waves of the  $\nu = \frac{1}{2}$
Laughlin state, which in the present context
corresponds to setting $p = n = 1$. 
However, a less straightforward exercise is provided by 
a state where $p = 2$ 
(and $\nu = \frac{2}{3}$ effectively). 
Using the parton construction, we find
that there are {\it two} independent branches of surface modes, 
described by  the currents 
$j_{0} = \left(2\rho_{0} + 
\rho_1 + \rho_2 \right)/3$, and $j_1 = \rho_1 - \rho_2$,  respectively,  
which obey 
\begin{eqnarray}
\left[ j_{0}(l), j_{0}(l') \right] &=& \frac{2}{3} l'\: \delta_{l+l',0},\\
\left[ j_{1}(l), j_{1}(l') \right] &=&  2 l'\:  \delta_{l+l',0}.
\end{eqnarray}
 As a check, it is worth trying 
to obtain these results from
a different point of view. Let us first recall that 
the $\nu = \frac{2}{3}$  state can be obtained
by means of the hierarchical construction~\cite{Haldane83} where 
the state at $\nu = 2/3$ can be 
thought of as two component vortex  liquid, 
containing one liquid  with $\nu_{0} 
= \frac{1}{2}$, which corresponds
to the {\it parent} Laughlin state,  and another one 
with $\nu_1 = \frac{1}{6}$, which corresponds to a
Laughlin state of quasi-particles over the parent state of bosons. 
Notice that the sum of filling
fractions $\nu_{0}  + \nu_{\rm 1} = \nu$. 
If, for the moment, we neglect  any interactions between
the two components of the liquid, 
the effective theory can be used to describe each one
separately. This  leads to the following current algebra:
\begin{eqnarray}
\left[ \sigma_{0}(l), \sigma_{0}(l') \right] = \frac{1}{2} l'\: \delta_{l+l',0},\\
\left[ \sigma_{1}(l), \sigma_{1}(l') \right] = \frac{1}{6} l'\: \delta_{l+l',0},
\end{eqnarray}
where $\sigma_{0}$ describes the boundary 
phonons of the $\nu_{0} = \frac{1}{2}$
liquid and $\sigma_1$ those of the $\nu_{1} 
= \frac{1}{6}$ liquid.

To find the relationship between 
the parton currents and the currents $\sigma_{0}$
and $\sigma_{1}$, we resort to
the interpretation of $j_{0}$ as the total current. This
implies that $j_{0} = \sigma_{0} 
+ \sigma_{1}$. The other current
must be an independent linear combination, which is
readily found  to  be $j_1 =  \sigma_{0} - 3 \sigma_1$. 
This, quite direct, identification of the currents helps to confirm 
the results obtained from the parton
construction. Thus we conclude that,
just as we briefly remarked in Sect.~\ref{sectlaughlin},
for states other than the Laughlin states there are
several branches of surface modes.
In particular, for the compact CF states considered above,
there can be $p > 1$ branches. 

  Let us finally turn our attention to  
the dynamic structure  factor for the states 
considered above. As it was pointed out previously,
these states are not exact ground states of the
Hamiltonian, which means that we cannot use
the wave function approach of Sect.~\ref{sectlaughlin}
to get much insight into the excitation energy of the
surface modes.  Furthermore, the non-interacting CF picture may only 
hold to describe the ground states but not their low-lying
excitations. In other words, some small residual interactions are always expected.
However, if we start by assuming that the non-interacting
CF holds, then the energy of the surface modes will 
be just the confinement energy, and therefore
the Hamiltonian reads:
\begin{equation}
H_{\rm o} = \hbar \pi \upsilon \sum_{\alpha=0}^{p-1}
\int_{0}^{2\pi} d\theta \:  J_{\alpha} ^2(\theta). 
\end{equation}
In the above expression we have normalized  the $p$ independent
currents so that $\left[ J_{\alpha}(l), 
J_{\beta}(l') \right] = l' \delta_{\alpha,\beta} \delta_{l+l',0}$ 
(notice that this normalization differs from the one used above for 
$j_0$ and $j_i$, $i=1,\ldots,n-1$). In the absence of interactions,
the  $p$ phonon branches are all  degenerate. However, 
since  the potential $\delta v_{LAB}(\theta)$ (see appendix~\ref{appa})
couples only to $j_0(\theta) = \sqrt{\nu} J_0(\theta)$ (where $\nu = p/(p+1)$),
$S_{LAB}(l,\omega)$ will have the form~(\ref{dsf}), displaying
a single peak at $\omega = \omega_{\perp} l$, but  with  $p/(p+1)$ replacing
$1/m$ or, in other words, with total spectral weight {\it proportional} 
to $p/(p+1)$. 

The  spectrum is modified, however, if the residual interactions could not be neglected.
In this case, one has to consider the effect of adding to the  Hamiltonian terms like:
\begin{equation}
H' = \hbar \pi \sum_{\alpha,\beta=0}^{p-1} \int_0^{2\pi} \Delta \upsilon_{\alpha \beta}\,
J_{\alpha}(\theta) J_{\beta}(\theta),
\end{equation}
where $\Delta \upsilon_{\alpha\beta}$ is a real 
symmetric matrix. One could also have included terms
containing derivatives of the currents, such 
like $ g_{\alpha \beta} \int d\theta \: J_{\alpha}(\theta) 
\partial_{\theta}J_{\beta}(\theta)$. Other terms containing
more than two current operators are also possible, and describe inelastic
scattering processes among the surface phonons. Being {\it irrelevant}
in the renormalization-group sense,  these terms
yield corrections to the energy of the surface modes
of the order of $ l (l/\sqrt{mN})^i$, where the index $i$ equals the order
of the current derivatives  plus the number of current operators minus two
(e.g. $i=1$ in the previous example containing two currents and one derivative).
These corrections can be important for modes with $l \sim \sqrt{N}$.  To  
the lowest order in this case, however, the ``normal" (surface) modes   can be 
obtained by  diagonalizing the matrix
\begin{equation} 
{\cal H}_{\alpha \beta} = \upsilon\:
 \delta_{\alpha\beta} + \Delta \upsilon_{\alpha \beta}.
\end{equation}
Nevertheless, the overall shape of the spectrum
is constrained by the generalized Kohn's theorem discussed
in appendix~\ref{appa}.  This theorem is a consequence of the
decoupling of the center of mass motion from other degrees of freedom  
in harmonically trapped systems. Since we have assumed that this
is the case throughout, it implies that the energy (in
the laboratory frame) of the $l=1$ mode
is given by the {\it bare} trap frequency $\omega_{\perp}$. 
Let us begin by considering the case where
there is a single phonon branch, like in the  
the Moore-Read state to be discussed below.  Kohn's theorem
implies that $\omega(l = 1) = \omega_{\perp}$. Any corrections
arising from interactions will  necessarily have to vanish for $l = 1$.
For instance, we can think of a general form for the dispersion (in the 
laboratory frame) like $\omega(l) = \omega_{\perp} l + 
l \sum_{i=1}^{+\infty}\delta \omega^{(i)}_{int} (l-1)^i$,
where $\delta \omega^{(i)}_{int} \ll \omega_{\perp}$. In the case where
there is more than one branch, the situation is slightly more
complicated.  Then, in principle, after diagonalizing the
matrix ${\cal H}_{\alpha\beta}$, not all the eigenvalues will
be degenerate, which implies that for $l = 1$, to this order,
there can be several modes with distinct energy. However,
the energy of, at least, one of the modes (the Kohn mode, associated with the
center of mass motion) is still fixed 
to $\omega_{\perp}$ (in the laboratory frame). Furthermore,
an analysis of the oscillator strength based on 
Kohn's theorem (see appendix~\ref{appa} and below) shows that the Kohn mode
exhausts the all the oscillator strength available, which in turn
forces all the modes to be degenerate at $l = 1$. Thus, the
general form of the dispersion of the {\it normal} modes is a generalization of
the single-branch case: $\omega(l,\alpha) = \omega_{\perp} l
+l \sum_{i=1}^{+\infty} \delta \omega_{int}^{(i)}(\alpha)  (l-1)^i$,
where $\alpha = 1, \ldots, p$ is the branch index. 
This form assumes the frequencies to be 
real and thus neglects any imaginary parts arising from inelastic
coupling of surface modes. However, this fact can be readily accounted for, although
we expect {\it this} linewidth to be small for the low-lying modes, and zero for  $l=1$. 

 The implications of the previous discussion for  the dynamic structure factor
can be easily extracted. After expressing the total current $j_0 = \sqrt{\nu} J_0$
in terms of the normal modes, we find that $S_{LAB}(l,\omega)$ 
should exhibit the following structure 
\begin{equation}
S_{LAB}(l,\omega) =  \sum_{\alpha = 1}^{p} \frac{l w_{\alpha} 
\delta(\omega - \omega(l,\alpha))}
{1-e^{-\hbar(\omega(l,\alpha) - \Omega l)/T}}.
\end{equation}
From the current algebra for $j_0(l)$, it follows the following sum rule:
\begin{equation}
\sum_{\alpha = 1}^{p} w_\alpha = \frac{p}{p+1}. 
\end{equation}
According to the above considerations, for $l = 1$  all the branches must 
be degenerate: $\omega(1,\alpha) = \omega_{\perp}$, and hence there is single peak 
of spectral weight proportional to $p/(p+1)$. 
For $l > 1$ more than one (but at most $p$) peak  can exist, and given that 
for repulsive interactions the energy of the Kohn 
mode is a lower bound of the surface phonon energy, all the peaks  should
appear at frequencies higher than (or equal to) $\omega_{\perp} l$. However, since
the interactions are weak, it may well happen that the shifts and splittings are very small
and hard to resolve experimentally, and this would produce a single peak again. 
The total spectral weight associated with the surface modes must be again proportional to
$p/(p+1)$, for integer $p \geq 1$. However, if the different peaks for $l > 1$ could be resolved, 
the individual spectral  weight of each peak  would be determined by the detailed 
form of the interactions between modes, which  
is not fixed by the present effective field theory.

 We close this section by considering  the surface modes of 
the Pfaffian or Moore-Read (MR)  state. This is a state that 
cannot be constructed from the
the composite-fermion ansatz, Eq.~(\ref{CF}).   As it was mentioned
at the beginning of the section, for $\nu= 1$ exact diagonalization studies
on the torus~\cite{Cooper01} and the sphere~\cite{Regnault02} 
have found an incompressible state exhibiting
a large overlap with the MR state,
whose  wave function reads:
 \begin{equation}\label{mr2}
 \Phi^{MR}_{B}(z_{1},\ldots, z_{N}) =
 {\cal A}\left[\frac{1}{z_{1}-z_{2}} \cdots
 \frac{1}{z_{N-1}-z_{N}} \right]
 \prod_{i<j=1}^{N} (z_{i}-z_{j}),
 \end{equation}
where ${\cal A} [\ldots]$ means that the bracketed 
product must be anti-symmetrized as follows ($N$ is assumed to 
be even):
\begin{equation}
{\cal A}\left[\frac{1}{z_{1}-z_{2}} \cdots
 \frac{1}{z_{N-1}-z_{N}} \right] = \frac{1}{2^{N/2} (N/2)!} \: \sum_{P \in S_N} (-1)^{P}\
 \frac{1}{z_{P(1)} - z_{P(2)}}\cdots \frac{1}{z_{P(N-1)}-z_{P(N)}}\: \: .
\end{equation}
In the previous expression $P$ is a permutation of $1,2,\ldots,N$ and $(-1)^P$ denotes its signature.
For instance, if we consider $N = 4$ particles, 
\begin{equation}
{\cal A}\left[\frac{1}{z_{1}-z_{2}}\: \frac{1}{z_{3}-z_{4}} \right] = 
\frac{1}{z_1-z_2}\: \frac{1}{z_3-z_4} - \frac{1}{z_1-z_3} \: \frac{1}{z_2-z_4} +
\frac{1}{z_1-z_4}\: \frac{1}{z_2-z_3}\: \:.
\end{equation}
Alternatively,  the MR state can be written in the following way~\cite{Cooper01,Wilkin00}:
\begin{equation}\label{MR}
\Phi^{MR}_{B}(z_{1},\ldots, z_{N}) = {\cal S} \left[ \prod_{(i < j) \in A} (z_i - z_j)^2 
\prod_{(k < l) \in B} (z_k - z_l)^2 \right],
\end{equation}
where ${\cal S}$ symmetrizes the product 
of the two Laughlin $\frac{1}{2}$ states over all possible 
partitions of $N$ particles into {\it two disjoint} subsets $A$ and $B$ having $N/2$ particles each.
The second form, Eq.~(\ref{mr2}),  shows perhaps more explicity 
that the lower angular momentum of the MR is attained by
having $N(N-1)/2 - N(N/2-1)/2 = (N/2)^2$ pairs 
of bosons in orbits with zero relative angular
momentum. This is to be contrasted with the Laughlin state, Eq.~(\ref{laughlin}),
where {\it all} boson pairs have relative angular momentum equal to $2\hbar$. 
This means that in the MR some particles must necessarily interact,  
in order to reduce the total angular momentum, which for this state equals
$N(N-2)/2$ for $N$ even and $L=(N-1)^{2}/2$ for $N$ odd.
Wilkin and Gunn found  in Ref.~\onlinecite{Wilkin00} that this state is also
a good candidate to describe the stable state found for rotating bosons 
in a harmonic trap at $L = N (N-2)/2$ (for even $N)$. 

Surprisingly, the MR state becomes an exact eigenstate with zero 
interaction energy for a model where the particles interact by means of 
a repulsive three-body potential ($g_3 > 0$)~\cite{Wen95}
\begin{equation}\label{three}
V_3 = g_3 \: \sum_{i<j<k} \delta^{(2)}({\bf r}_i - {\bf r}_k) \delta^{(2)}({\bf r}_i-{\bf r}_j).
\end{equation}
This follows from the property that MR vanishes when any three-particles
come together, and has implications for the stability of the state with respect to
three-body losses. We have previously found that the Laughlin state was a zero energy
eigenstate of a two-body interaction potential. This is because its
wave function vanishes as any two-particles approach each other, hence the probability
for three particles to be at the same point must be necessarily zero as well. 
Therefore,  for the Laughlin and the MR states the
three-body recombination rate~\cite{Kagan88} should be zero, rendering them very stable. However,
the MR state is only a good approximation to the true ground state of a boson system
interacting with two-body potentials, which implies that in the actual system the recombination rate
should be small but not zero.

	However, the existence of a Hamiltonian for which the MR state has zero energy
allows to perform a {\it microscopic} analysis of the surface excitations formally 
analogous to the one carried out for the Laughlin state. 
Thus it is found that the symmetric polynomials $s_l = \sum_{i=1}^{N} z^l_i$
describe, also in this case, low-lying excited states of the Hamiltonian $V_3$ introduced above:
The states obtained multiplying the MR state by $s_l$ have zero interaction energy, and
therefore are degenerate with the MR state,  but  adding the confinement 
term, $\upsilon L$, this degeneracy is lifted. We can identify the surface modes 
of the MR state with the polynomials $s_l$, exactly as it was 
done for the Laughlin state.  But it has been
shown by several authors~\cite{Wen95,Milovanovic96} that these excitations do not exhaust
the spectrum of boundary modes of the MR state. Besides the chiral phonons, which have 
bosonic character,  fermionic excitations that do not carry charge quantum numbers (i.e. they
are described by   a Majorana fermion) also exist.  Lack of charge quantum numbers means that it
is not expected that these excitations will  be {\it directly} created 
by weakly  deforming  the confining potential (it can happen that the fermionic 
neutral  and the bosonic charged excitations
are somehow coupled, but the form of the interaction seems 
not to be easy to write~\cite{huertapc}). 
Thus we will focus only on the single branch of ``charged'' 
surface phonons. By the same arguments employed with other states of scalar bosons, the phonon
energy  is given (in the laboratory frame) by $\omega = \omega_{\perp} l$. But 
since the  MR state is indeed an approximation, we cannot exclude the possibility that the phonon dispersion 
will receive corrections  for $l > 1$ (for $l=1$ the energy is fixed by Kohn's theorem,
as discussed above). However, this does 
not affect our conclusion that, for the MR state, the dynamic structure factor
should display a single surface phonon peak, whose energy is approximately linear with 
the angular momentum of the mode. One can repeat the steps that lead to the effective field 
theory  for a droplet of MR  liquid, and obtain that the spectral weight of the 
surface phonon in $S_{LAB}(l,\omega)$ is  again proportional to $1/m$ with $m=1$.


\section{Vector Bosons}\label{vecbosons}
 
 Let us now take up bosons with internal (i.e. hyperfine) degrees of freedom. 
We call them vector bosons because we will describe them by 
a  field operator that  is a tuple of $n$-fields,
\begin{equation}\label{vecfieldop}
\vec{\Psi}^{\dag}(z) = (\Psi^{\dag}_0(z), \ldots, 
\Psi^{\dag}_{n-1}(z)),
\end{equation}
and which transforms as a vector under $SU(n)$ transformations. This does not
necessarily mean that the Hamiltonian has this symmetry, but we will find that 
under certain conditions the effective theory describing the surface modes  does. 
One can also regard the internal 
degree of freedom as  a spin or pseudo-spin, depending on the context.  Thus
an $n$-component vector boson corresponds to a boson of 
spin $S= (n-1)/2$.  Again, this does not mean
that the Hamiltonian is in general rotationally invariant (see below). 
Scalar bosons (i.e. $S=0$) are obtained by letting $n=1$. For $n = 2$ we obtain
spin-$1/2$ bosons, which can be used to describe a mixture of two hyperfine states
of the same isotope (as in the experiment reported in Ref.~\onlinecite{Myatt97}). 
If the atoms are trapped by purely optical means, one 
can have in the same trap~\cite{Ketterle00} the three magnetic
sublevels (i.e. $m_F = -1, 0, +1$) of an isotope of total spin $F = 1$~\footnote{In what follows,
we shall also denote by $S$ the hyperfine spin $F$.}, and this 
case will be treated by setting $n = 3$.  One also  can imagine experiments with atoms of  higher $S$ ($F$). 
In general, however,  the interaction 
will depend on the hyperfine state of the
colliding atoms. Thus for $S = 1/2$, it can take  the generic form
\begin{equation}
V_{ij} = \left[ g_0 + g_1 ( \Delta_x S^{x}_i S^{x}_j + \Delta_y S^{y}_i S^{y}_j +  \Delta_z S^{z}_i S^{z}_j)\right] 
\delta({\bf r}_i - {\bf r}_j), 
\end{equation}
for any two particles $i$ and $j$. For spin-$1$ bosons in an optical trap 
rotational invariance implies that~\cite{Pethick02}
\begin{equation}
V_{ij} = \left[ g_0 + g_1 {\bf S}_i \cdot {\bf S}_j\right] 
\delta({\bf r}_i - {\bf r}_j), 
\end{equation}
Higher spin will involve higher powers of  $({\bf S}_i\cdot {\bf S}_j)$~\cite{Pethick02}.
Typically one has  $|g_1| \ll g_0$, while the sign of $g_1$, which determines the ferromagnetic 
(i.e. $g_1 <0$, like in $^{87}$Rb) or anti-ferromagnetic (i.e. $g_1 > 0$, like $^{23}$Na) character of
the spin-interaction, depends on the atom species. 

	We have seen in previous sections that as the amount of angular momentum deposited in 
the system is increased, all atoms tend to avoid each other. When the atoms
have internal degrees of freedom, a cloud containing an equal number of atoms
in each internal state will go to a singlet state with zero interaction energy at sufficiently high
angular momentum.  This possibility has been recently considered by a number of authors for spin-$1$~\cite{Ho02,Reijnders02} as well as for arbitrary spin bosons~\cite{Paredes02}. 
The parton  construction can be readily applied to study these states. We will consider
that the interaction is dominated by the term proportional to $g_0$, but including a
weak (as it is the usually the case) spin-dependent scattering term makes no difference. 
We stress that this is because, as long as the interaction has zero-range, 
these states have zero interaction energy because particles completely avoid each other
independently of their internal state.

To see how the parton construction allows us to describe singlet states, we begin by writing
the components of field operator, Eq.~(\ref{vecfieldop}),  as the product of two 
field operators of the fermionic partons, i.e.
\begin{equation}\label{fieldopCF}
\Psi^{\dagger}_{\alpha}(z) = \psi^{\dagger}_{1}(z) \psi^{\dagger}_{2\alpha}(z),
\end{equation}
for $\alpha = 0, \ldots, n-1$.  In terms of wave functions 
this implies that the boson wave function reads
\begin{equation}\label{vecCF}
\Phi_{B}(z_1 \alpha_1, \ldots, z_N \alpha_N) = 
\Phi_{1}(z^{1}_1, \ldots, z^{1}_N)
\Phi^{(n)}(z^{2}_1 \alpha_1, \ldots, z^{2}_N \alpha_N)\Big|_{ \left\{ { z^{1}_i 
= z^{2}_i  }\right\}_{i=1,\ldots,N}},
\end{equation}
where $\Phi_{1} = \prod_{i<j} (z^{1}_i - z^{1}_j)$ denotes the 
Jastrow factor, which describes the state of type-$1$ partons,
whereas $\Phi^{(n)}$ is a Slater
determinant of $N$ type-$2$ partons in 
internal states $\alpha_1,\ldots, \alpha_N$. 
In this context we are interested in states where all 
type-$2$ partons are in the LLL. The situation differs
from the case of scalar bosons because now type-$2$ partons
carry the internal  degrees of freedom, and therefore
states other than the Laughlin state will appear
for which the projection onto the LLL is unnecessary.  It is interesting
to point out  that it is the Jastrow factor that  ensures that 
the resulting wave function has zero interaction
energy, while the kinetic energy is quenched to $N \hbar
\omega_{\perp}$ plus the confinement energy. In  other words, the
factor $\Phi_{1}$ makes the wave 
function vanish whenever {\it any} two-particles
approach each other, independently of their internal quantum state. 
For the same reason, the probability for three particles
to approach each other is zero, which in turn implies
that the three-body recombination rate vanishes: These
states should be particularly long-lived.

To make  things more concrete, we specialize our discussion
to  $n =  2$ components, i.e. $S = \frac{1}{2}$
bosons.  We consider this case first because it is simpler but 
already illustrates the important points, and also because it is relevant to
the situation where only two hyperfine states are allowed in the same trap~\cite{Myatt97}. 
We will discuss the generalization to arbitrary 
$n$ at the end of this section. Setting $n = 2$, we assume that  $\alpha = 0$ 
corresponds to spin up  ($\uparrow$) and $\alpha = 1$ to  spin down ($\downarrow$).
As mentioned previously, we shall consider only singlet  ground states. 
Therefore, the number of particles, $N$, must be
divisible by $n = 2$, i.e. $N_{\uparrow} = N_{\downarrow} 
= N/2$ . Nevertheless, the latter condition
does not suffice  for $\Phi_{B}$ 
to be a singlet but only ensures 
that $S_z |\Phi_{B} \rangle = 0$. 
In addition it is needed that
\begin{equation}\label{singlet}
S^{-} |\Phi_{B} \rangle = 0,
\end{equation}
where $S^{-} = \sum_{j=1}^{N} S^{-}_{j}  
= \sum_{j=1}^{N} (S_x - i S_y)$ is the operator
that decreases the total spin.
As the spin index is carried by  type-$2$ partons, the above 
condition must  be  indeed met by the parton 
wave function $\Phi^{(2)}$. Upon filling the LLL with $N/2$ fermions
of both spin species, the parton wave function takes the form
\begin{equation}\label{partonsinglet}
|\Phi^{(2)} \rangle = \prod_{l=0}^{N/2-1} \psi^{\dagger}_{2\uparrow}(l)
  \psi^{\dagger}_{2\downarrow}(l) |0\rangle,
\end{equation}
where $|0\rangle$ is the zero particle state. 
It is not hard to see that this state 
obeys the condition~(\ref{singlet}) 
since the operator $S^{-} = \sum_{l=0}^{+\infty}
 \psi^{\dagger}_{2\downarrow}(l)\psi_{2\uparrow}(l)$
annihilates it. However, in order to obtain the boson wave function, 
we need the spatial dependence of~(\ref{partonsinglet}). 
This is just the product of two
Slater determinants of $N/2$ fermions in the LLL, 
one for each spin orientation, which have the form of the
Jastrow wave functions that we have been using throughout. 
Hence, using Eq.~(\ref{vecCF}), one finds 
\begin{equation}\label{Halperin}
\Phi_{221}(z_1,\ldots, z_{N/2}; w_1, \ldots, w_{N/2}) = 
\prod_{i<j=1}^{N/2} (z_i - z_j)^{2} (w_i-w_j)^2 \: 
\prod_{i,j=1}^{N/2}(z_i-w_j),
\end{equation}
where $z_i$ and $w_i$ ($i=1,\ldots, N/2$) 
denote the positions of the spin up and down bosons,
respectively. Notice that this wave function seems  
to have a rather peculiar dependence 
on $z_i$ and $w_i$, which is not fully 
symmetric in these variables,  in spite of being
a bosonic wave function. However, 
one must be careful since we are dealing with
spinful bosons. Indeed, the above form is what the
singlet condition requires. As the reader can easily 
check  for  two bosons making a singlet,
the anti-symmetry of the spin part 
of the wave function  requires 
an anti-symmetric spatial wave function. 
The appropriate generalization of this observation
to many particles is given by~(\ref{Halperin}). 
The $\Phi_{221}$ wave function is indeed
one of the class introduced  
by Halperin~\cite{Halperin83,Yoshioka02} 
to describe non-fully spin-polarized 
quantum Hall states.
In Ref.~\onlinecite{Reijnders02} this
state was characterized by the {\it affine} 
algebra $su(3)_1$, i.e 
the symmetry of the two-dimensional (conformal)
field theory  for which the wave function can be obtained
as a correlation function (see  the appendix for a simple
example, and below). 

 After explaining how to construct  some of the vector boson states  
from partons, we  proceed with the description 
of their surface modes. Indeed, all we need 
to do is to re-interpret what
was done in the previous section for the $n = p =  2$ state. 
Now $j_{0}$ must be interpreted
as the charge current, which we  denote  as $j_c$ in 
what follows, whereas  $j_1/2$ measures the density of  
the third component of the spin at the boundary, 
and we denote it as $j^{3}_{s}$. Therefore,
\begin{eqnarray}
j_{c}(\theta) &=& \frac{2}{3} \left[\rho_{0}(\theta) 
+ \frac{1}{2} \left(\rho_{\uparrow} + \rho_{\downarrow} \right)(\theta)  \right],\\
j^{3}_{s}(\theta)  &=&\frac{1}{2} \left( \rho_{\uparrow} - \rho_{\downarrow} \right)(\theta) = \frac{1}{2} \left[
\psi_{2\uparrow}^{\dagger}(\theta) 
\psi_{2\uparrow}(\theta) - \psi_{2\downarrow}^{\dagger}(\theta) \psi_{2\downarrow}(\theta) \right].
\end{eqnarray}
A difference with respect to the case of scalar bosons is that, in this case,
$j_c$ and $j^{3}_s$ are decoupled, as long as the trapping potential is spin-independent. 
The symmetry of the problem is $U(1)\times SU(2)$, where
where  $U(1)$  is associated with the total 
particle number (i.e. the charge), and therefore with $j_c$,
whereas $SU(2)$ is related to the internal degree of freedom, and
therefore  to the spin current, $j^3_s$. Indeed, 
the spin current is part of the generators of
a  larger algebra, which also includes the  currents 
\begin{eqnarray}
j^{1}_{s} &=&  \frac{1}{2} \left[ \psi^{\dagger}_{2\uparrow} \psi_{2\downarrow} + 
\psi^{\dagger}_{2\downarrow} \psi_{2\uparrow} \right], \\
j^{2}_{s} &=& \frac{1}{2i} \left[ \psi^{\dagger}_{2\uparrow} \psi_{2\downarrow} - 
\psi^{\dagger}_{2\downarrow} \psi_{2\uparrow} \right].
\end{eqnarray}
 These three currents can be expressed 
in a more compact way as $j^{a}_s(\theta) = \frac{1}{2}\sum_{\alpha,\beta}
\psi^{\dagger}_{\alpha}(\theta)
\sigma^{a}_{\alpha\beta} \psi_{\beta}(\theta)$,
and they obey the following  algebra ($a,b, c = 1, 2, 3$):
\begin{equation}
\left[ j^{a}_{s}(l), j^{b}_{s}(l') \right]
=  \frac{l'}{2} \delta^{ab}\delta_{l+l',0} +
 i\epsilon^{ab}_{c} \: j^{c}_c(l+l'),
\end{equation}
known as $su(2)_1$ Kac-Moody algebra.

  If decoupled,
the charge, $j_{c}$, and spin, $j^{3}_{s}$,
currents describe waves that propagate
with the same frequency $\upsilon$. This can be shown
using the wave function approach introduced in Sect.~\ref{sectlaughlin}.
In this case, each spin orientation is related to different set
of symmetric polynomials, $t_l = \sum_{i=1}^{N/2} z^l_i$,
$u_l = \sum_{i=1}^{N/2} w^l_i$, with the same excitation energy,
$\hbar \upsilon l$. The charge modes corresponding
to the combination $t_l + u_l$ and the spin modes to
$t_l - u_l$, are degenerate. Thus we conclude that 
the dynamics of these modes is dictated by the confining potential
(which we have assumed spin-independent). Just like for the Laughlin state,
this is a consequence  of the wave function  $\Phi_{221}$ being
an exact eigenstate of the Hamiltonian with zero 
interaction energy.  As far as the observable consequences are concerned,
it is now possible to consider  two dynamic structure factors,
given by the correlation functions $\langle j_c(\theta,t) j_c(0)\rangle$
and $\langle j^3_s(\theta,t) j^3_s(0) \rangle$, and related to  
deformation potentials that couple to the total density, $j_c$,
or the spin density, $j^3_s$. In both cases, the observable
spectrum will exhibit a single peak at $\omega = \omega_{\perp} l$.
The difference will be in the spectral weight of the peaks,
which for the charge peak will be proportional to $\nu = 2/3$,
whereas the spin peak to $1/2$. We finally point out that
if the confining potential is spin-dependent, then the separation
into a charge and a spin density current is no longer convenient. In other words,
the currents $j_c$ and $j^2_s$ do not correspond to the normal
modes of the system any more. In such a case, for instance, 
the spectrum for a trapping potential deformation 
that couples to the total charge  will display two peaks at
two different frequencies $\omega = \omega_{\uparrow\perp} l$
and $\omega = \omega_{\downarrow\perp} l$ (we assume that 
$\omega_{\uparrow,\downarrow\perp} - \Omega\ll \Omega$). The total
spectral weight of the two peaks, however, will be 
again proportional to $2/3$, each peak contributing $1/3$.

	As a final demonstration of the internal consistency of 
these constructions, we will rederive the
wave function $\Phi_{221}$ using the
formalism explained in the appendix, and which relates
the wave function to the correlation functions of the
effective theory for surface modes. 
The key to this approach is to identify the boson operator
in the LLL with the vertex operator (see appendix for a definition)
that describes the boson operator at the boundary. To this purpose, 
let us introduce the  phonon fields for the partons, in analogy to
how it was defined in Sect.~\ref{sectlaughlin}, i.e. 
\begin{eqnarray}
\partial_{\theta}\phi_{0}(\theta) &=&  2\pi
\rho_{0}(\theta) ,\\
\partial_{\theta}\phi_{\alpha}(\theta) &=& 2\pi
\rho_{\alpha}(\theta),  
\end{eqnarray}
with $\alpha=\uparrow, \downarrow$. The the parton
construction dictates that the boson
field operator at the boundary takes the form~\cite{Wen92}
\begin{equation}
\Psi^\dag_{\alpha}(\theta) =
 e^{-i\left(\phi_{0} + 
 \phi_{\alpha} \right)(\theta)},
\end{equation}
which is nothing but Eq.~(\ref{fieldopCF}) with the parton fields written as
exponentials of their corresponding phonon fields (cf. Sect.~\ref{sectlaughlin}).
However, the phonon fields $\phi_{0}, \phi_{\uparrow}, \phi_{\downarrow}$,
just as the currents $\rho_{0},\rho_{\uparrow}, \rho_{\downarrow}$,  are not independent. Only
$j_{c}$ and $j^3_{s}$ are independent.  But the following expressions:
\begin{eqnarray}
\rho_{0} + \rho_{\uparrow} &=& \frac{3}{2} j_{c} 
+  j^3_{s}, \\
\rho_{0} + \rho_{\downarrow} &=& \frac{3}{2} j_{c} 
-  j^3_{s},
\end{eqnarray}
help us to relate the phonon field combinations 
that enter in $\Psi^{\dag}_{\alpha}$ to the charge and
spin phonon fields, defined by the following equations:
\begin{eqnarray}
\partial_{\theta}\varphi_{c}(\theta) &=&
2\pi\sqrt{\frac{3}{2}}\: j_{c}(\theta),\\
\partial_{\theta}\varphi_{s}(\theta) &=&
2 \pi \sqrt{2}  \: j^3_{s}(\theta ).
\end{eqnarray}
These fields have been so normalized  that their
vertex operators  have the 
two-point correlation functions:
\begin{equation}
\langle V^{i}_{+1}(\bar{z}) V^j_{-1}(0)
\rangle=
\langle :e^{i\varphi_{i}(\bar{z})}: \:
:e^{-i\varphi_{j}(0)}:\rangle = \frac{\delta_{ij}}{z},
\end{equation}
where $i,j=c,s$. The field operators read
\begin{eqnarray}
\Psi^{\dagger}_{B\uparrow}(\bar{z}) &=& \, :e^{-i\left(\sqrt{\frac{3}{2}}\varphi_{c} 
+ \frac{1}{\sqrt{2}}\varphi_{s}\right)(\bar{z})}:\,\, , \\
\Psi^{\dagger}_{B\downarrow}(\bar{z}) &=& \, :e^{-i\left(\sqrt{\frac{3}{2}}\varphi_{c} - \frac{1}{\sqrt{2}}\varphi_{s}\right)(\bar{z})}:\, \, .
\end{eqnarray}
We are now ready to compute the wave function.
It takes the form
\begin{eqnarray}
&\Phi_{221}&(\bar{z}_1,\ldots,\bar{z}_{N/2};
\bar{w}_{1}, \ldots, \bar{w}_{N/2}) =
\langle \prod_{i=1}^{N/2}
\Psi^{\dag}_{\uparrow}(\bar{z}_i) 
\Psi^{\dag}_{\uparrow}(\bar{w}_i)
\: :e^{i\sqrt{\frac{3}{2}} \int\limits_{|z|<R} 
d^2z \, \varphi_{c}(\bar{z})\rho_{\rm o} }:\rangle
\nonumber\\ 
&=& \langle \prod_{i=1}^{N/2} V^{s}_{-\sqrt{\frac{1}{2}}}(\bar{z}_{i})
V^{s}_{+\sqrt{\frac{1}{2}}}(\bar{w}_{i})\rangle\,
\langle \prod_{i=1}^{N/2} V^{c}_{-\sqrt{\frac{3}{2}}}(\bar{z}_{i})
V^{c}_{-\sqrt{\frac{3}{2}}}(\bar{w}_{i})
\: :e^{i\sqrt{\frac{3}{2}} \int\limits_{|z|<R} 
d^2 z\, \varphi_{c}(\bar{z})\rho_{\rm o} }:
\rangle  \nonumber \\
&=&  \prod_{i<j=1}^{N/2}
(\bar{z}_{i}-\bar{z}_{j})^{1/2} 
(\bar{w}_{i}-\bar{w}_{j})^{1/2} 
\prod_{i,j=1}^{N/2} 
(\bar{z}_{i}-\bar{w}_{j})^{-1/2} \nonumber\\
&&\times
\prod_{i<j=1}^{N/2} (\bar{z}_{i}-\bar{z}_{j})^{3/2}
 (\bar{w}_{i}-\bar{w}_{j})^{3/2} \prod_{i,j=1}^{N/2}
 (\bar{z}_{i}-\bar{w}_{j})^{3/2} \prod_{i=1}^{N/2} 
 e^{-(|z_{i}|^2 + |w_{i}|^2)/2\ell^2} \nonumber\\
&=& \prod_{i<j=1}^{N/2} (\bar{z}_{i}-\bar{z}_{j})^2
(\bar{w}_{i}-\bar{w}_{j})^2 
\prod_{i,j=1}^{N/2} (\bar{z}_{i}-\bar{w}_{j}) \, 
 \prod_{i=1}^{N/2} 
 e^{-(|z_{i}|^2 + |w_{i}|^2)/2\ell^2}.
\end{eqnarray}
This is just the anti-analytic version of the 
Halperin 221 state, Eq.~(\ref{Halperin}). In the above
expressions $R$ is the radius of the disk of positive charge
background with  density 
$\rho_{\rm o} = \nu/\pi \ell^2 = 2/(3\pi \ell^2)$
and total charge equal to $N$. It is also interesting to notice that,
in the above construction the wave function splits into two 
factors, one involving only charge ($\varphi_c$)  and
other involving only spin ($\varphi_s$) operators. This decomposition
can be understood as due to the separation of the fundamental boson
into two kinds of excitations: a {\it spinon}, which carries the spin, and a {\it holon}, which
carries  the ``charge''.

 Finally, the generalization to arbitrary $n$ is not hard
to work out.  For a number of particles divisible
by $n$, the singlet $SU(n)$  wave function reads:
\begin{equation}
\Phi\left({\left\{ z^{\alpha}_1,\ldots, z^{\alpha}_{N/n}\right\}}_{\alpha=0,\ldots,n-1}\right)
= \left[ \prod_{\alpha=0}^{n-1}\prod_{i<j=1}^{N/n} (z^{\alpha}_i - z^{\alpha}_j)^2 \right]
\left[ \prod_{\alpha < \beta} \prod_{i,j=1}^{N/n} (z^{\alpha}_i - z^{\beta}_j) \right],
\end{equation}
where $\alpha, \beta = 0, \ldots, n-1$ (This state is the $su(n+1)_1$ state of Ref.~\cite{Reijnders02}).
Just as we did above, one can construct the surface modes from the symmetric polynomials,
$s^{(\alpha)}_l = \sum_{i=1}^{N/n} \left( z^{(\alpha)}_i\right)^l$, which will have excitation energy 
(in the rotating frame) $\hbar \upsilon_{\alpha} l$, with  $\upsilon_{0} = \upsilon_1 =
\ldots =  \upsilon_{n-1} = \upsilon \equiv \omega_{\perp} - \Omega$ if the trapping potential is spin-independent.
The parton construction yields $n$ independent currents, one of them,
$j_c = (n \rho_0 + \sum_{i=1}^{n-1} \rho_{i})/(n+1)$,  corresponding to the
total current and is therefore associated with the  charge $U(1)$ symmetry.
The remaining $n-1$ currents are part of a $su(n)_1$ algebra that describes the
dynamics of the internal degree of freedom.  The dynamic structure factor 
related to the $\langle j_c(\theta,t) j_c(0)\rangle$ 
correlator exhibits a single peak at $\omega = \omega_{\perp} l$, whose intensity is
proportional to $\nu = n/(n+1)$. If the trapping potential is spin-dependent, however,
this peak splits into as many peaks as distinct values of the frequency $\omega_{\alpha\perp}$ exist,
but the total intensity remains proportional to $n/(n+1)$ each spin component contributing $1/(n+1)$.

         
\section{Conclusions and Outlook}\label{concl}

In this paper we have described the surface modes of  
ultra-cold atomic clouds with a very large number of vortices. These
systems are known as vortex liquids, and
are expected to be observed 
in the critical rotation regime where
the rotation frequency approaches the trap frequency. 
The surface waves of vortex liquids present several important differences with
respect to the surface modes of  a Bose condensate. For an axially symmetric
trap of the type considered above, the frequency dispersion of the 
surface modes of a BEC is found to be~\cite{Pethick02}:
\begin{equation}
\omega = \omega_{\perp} \sqrt{|l|},
\end{equation}
where $l = 0, \pm 1, \pm 2, \ldots $ On the other hand, for a spin-independent
trap,  a vortex liquid exhibits  surface modes whose dispersion 
is well approximated by 
\begin{equation}\label{w}
\omega = \omega_{\perp}  l,
\end{equation}
where $l = 1, 2, 3, \ldots$ i.e. a positive integer, which measures by how many 
quanta is increased the total angular momentum of the vortex liquid. We also found
other (approximate) excitations, which involve promoting particles to higher
Landau levels, and which have the same energy as the surface modes in the laboratory
frame. However, these excitations decrease the total angular momentum 
by $l = 1, 2, \ldots $ quanta. However, we must stress that the
surface modes are the true low-lying excitations of the vortex liquids as they 
dominate their low-temperature properties (for $T \ll \hbar \Omega$) since it is their energy
in the rotating frame, not that in the laboratory frame, that enters the Boltzmann factor. 
Furthermore, we argue that for $T \gg \hbar (\omega_{\perp} - \Omega)$,
their intensity in the dynamic structure factor should be much larger.  

 The chirality of the surface modes, namely that
$l$ is restricted to  positive integers, is
a consequence of the dramatic effect that rapid rotation  has on
a cloud of ultra-cold atoms: Besides enhancing quantum fluctuations, which are
finally responsible for the destruction of the condensate and the melting
of the Abrikosov lattice, fast rotation, when regarded from the rotating reference
frame, has the same effect as a strong magnetic field 
on a system of  charged particles moving in two dimensions. In both situations,
time-reversal symmetry is broken, which leads to the chirality of surface modes. 
We find, by comparing with the angular momentum and excitation energy
of  bulk excitations of  the Laughlin liquid, that the 
surface waves are the low-lying modes of the system. Indeed,
quasi-holes and quasi-particles are bulk excitations that change
the total angular momentum by $\sim \hbar N$, whereas the surface modes
do it just by $l \lesssim \sqrt{mN}$ quanta. We expect that this property also holds for the more 
complicated vortex liquids described in Sect.~\ref{other}. For these states, we
find that there exist more than one branch of surface phonons. This result 
is obtained by relying on the composite-fermion ansatz used 
in Ref.~\onlinecite{Cooper99} and first introduced for 
fermions in Ref.~\onlinecite{Jain95}. In this picture, a bound 
state of a boson with a vortex is an object of fermionic statistics,
known as composite fermion, which occupies $n$  ``renormalized''
Landau levels. The parton construction~\cite{Wen92} used in
our calculations in a concrete implementation   to study  surface excitations
of the composite-fermion idea.

	We  propose to probe the surface excitations  by adding a
weak time-dependent part  to the trapping potential that imparts small amounts
of angular momentum to the system. This should excite
surface waves in droplets of vortex liquid. We have quantified the response of the system
in terms of the dynamic structure factor $S_{LAB}(l,\omega)$. 
This function is related to the density 
correlation-function at the boundary.  If the residual interactions between 
the composite fermions can be neglected, $S_{LAB}(l,\omega)$ should exhibit
a single peak at the frequency given by Eq.~(\ref{w}). However, in the case where
residual interactions are not negligible,  several peaks may appear for $l > 1$. 
We also find that the spectral weight of the surface modes in $S_{LAB}(l,\omega)$, 
is proportional  to $p/(p+1)$, where $p$ is a positive integer. This means that if the system is gradually
driven to states of higher angular momentum, going through different vortex liquids  
at every stage, the integer $p$ characterizing the peak intensity of the surface modes, 
should decrease down to $p=1$, which corresponds to the system being in Laughlin state with 
$\nu = \frac{1}{2}$ (higher angular momentum states can be imagined but this would
be experimentally much harder).  In other words, the total intensity in 
$S_{LAB}(l,\omega)$ of the surface modes
at a given $l$  should decrease as  the Laughlin $\frac{1}{2}$ state is approached, at
constant temperature.

	Finally, we have also found chiral surface waves in vortex liquids
of vector (i.e. spinful) bosons. In this case, the parton construction, used to study
the surface waves of scalar bosons, can be also successfully applied. If the trapping potential  is spin
independent, the experimental signatures are similar to those for states of scalar bosons
with $p$ equal to the number of components of the boson field ($2$ for spin $\frac{1}{2}$ bosons,
$3$ for $S = 1$, ..., $2S +1$ spin-$S$ bosons). However, in this case the 
surface-phonon peak is always to 
be found for $\omega = \omega_{\perp}  l$, where $l > 0$. In 
spin-dependent traps, as long as the difference $\upsilon_{\alpha} = \omega_{\alpha\perp} - 
\Omega \ll \Omega$ ($\alpha =1,\ldots, 2S + 1$), the peak  splits into 
as many peaks as distinct values of $\omega_{\alpha\perp}$ exist.

	The study of the surface modes has also produced other results. We have found
that  the peak intensity of the surface phonon in the Laughlin liquid is proportional to $1/m$, where
$m$ is the inverse of the filling fraction. This was interpreted as if the bosons
in the Laughlin liquid behave as ``super-fermions'' which, on average, occupy two
states per particle.  This is a particular example of the generalized exclusion statistics  first
discussed by Haldane~\cite{Haldane91}. This feature of the spectral
weight of the surface modes of the Laughlin state can be interpreted as  
an experimental manifestation  of this exotic exclusion statistics. In addition, 
by analyzing the behavior of 
one-body density matrix near the boundary, we are able
to obtain the occupancy of the orbitals with the highest angular momentum in the Laughlin state.
For $m > 1$, we have argued that the behavior of the occupancy near the boundary forces
a maximum in the occupancy before it decays to zero at $l = l_{max} = m(N-1)$, where $N$ is
the number of particles in the droplet. This maximum has its correspondence in the density
profile, which should be a hallmark of the Laughlin liquid, and possibly of other vortex liquids as well.
Last but not least, we have argued that the Laughlin liquid and the singlet states of vector
bosons, should be very stable with respect to three-body losses. This is because their
wave functions vanish when any two particles approach each other. The Moore-Read 
state that provides a good approximation to  the state of total angular 
momentum $L \approx N^2/2$ for scalar bosons, should also be very
stable since its wave function vanishes when any three particles come to the same point.
However, since this state is only a good approximation to the actual ground state, we expect it 
to be less stable than the Laughlin state, which is exact.

  In spite of the successes of the constructions presented in this work, there remain a number
of important open problems. An important one is related to the accuracy of the description given
here to describe the surface modes of vortex liquids of scalar bosons. The description seems to
be well established for the Laughlin state, but for the other stable states found at lower angular momentum
it should be checked numerically. This can be done by analyzing the low-lying excitations of the
states found in Ref.~\onlinecite{Cooper99}. We expect that the analysis performed  here 
will encourage further studies in this direction, which may prove or disprove the conclusions
of our analytical work. Another interesting issue that has not been addressed in this article is 
the possibility that not all the excitations predicted by the parton construction are indeed surface
modes. This problem was  numerically analyzed for fermion systems in Ref.~\onlinecite{Cappelli98},
where it was argued that the effective theory for the surface waves of states in the (fermionic) Jain 
sequence is a ${\cal W}_{1+\infty}$ minimal model instead of the larger (in the sense
that it has larger spectral degeneracies, but the same spectrum) $u(1) \times su(n)_1$ model, which 
results from the parton construction used in the present work (see Ref.~\onlinecite{Cappelli98} and
references therein). Similar comments  apply to the other states of vector bosons which 
have not been considered in this paper, but which were recently constructed 
in Ref.~\cite{Reijnders02}. To sum up,
we can say that we have just begun to scratch the surface of the rich 
phenomena  offered by this new setup 
for strong correlation and quantum Hall physics.


\acknowledgments

 The author wants to express his gratitude to Erich Mueller for
a useful suggestion. He also gratefully  acknowledges 
useful correspondence with Nicola Wilkin and helpful 
conversations with Ignacio Cirac, Marina Huerta, Alexander Nersesyan,
Antonio Sarsa, Subodh Shenoy, and Peter Zoller. Financial 
support provided by the ESF programme ``BEC 2000+'' for a visit to the MPQ in
Garching (Germany) is  gratefully acknowledged.

\appendix 

\section{Relationship between properties in the laboratory and rotating frames. The
dipole mode and generalized Kohn's theorem}\label{appa}

   In BEC's of ultra-cold dilute gases  time-of-flight and {\it in situ} imaging of shape
oscillations, as well as Bragg scattering~\cite{Stamper99,Ketterle00}, have been used
to measure the excitation spectrum. Light scattering in general, and Bragg scattering
in particular, allow  to access the dynamic structure factor $S({\bf q},\omega)$~\cite{Stamper99,Ketterle00}.  
Whatever the method of choice,  here we shall imagine a situation where
a time-dependent non axi-symmetric potential  is added to the  trapping potential in
such a way that it drives the system slightly out of equilibrium by exciting surface modes. 
The response to the potential is measured and since
we are interested in the  {\it linear response} regime, we shall assume throughout that
the perturbation is weak.  All the measurements are performed 
in the laboratory (or {\sc lab}, for short) reference  frame. At this point
it is convenient to recall the steps needed in the preparation of a rotating cloud. 
One of the commonly used methods  consists in adding a time-dependent and non-cylindrically
symmetric part, pretty much like the perturbation that we shall consider later, which imparts 
angular momentum to a conveniently cooled BEC. We shall assume that the cloud has been subjected to this
``stirring potential'' for some time, driven at some angular frequency $\Omega_d$. When enough
angular momentum has been deposited in the system, the external force is turned off and the 
could reverts to a cylindrically symmetric situation, where the  Hamiltonian 
has the form discussed in Sect.~\ref{disgr}. Under these conditions, for an axially symmetric
trap, the angular momentum along the axis, $L$, is a  conserved quantity (i.e. it commutes with
the Hamiltonian). This means that the total angular momentum imparted to the system will remain
constant subsequently, rotating at a frequency $\Omega \neq \Omega_d$~\cite{Ho01}.
Next we imagine that in this situation the system is perturbed again by switching
one a weak time-dependent stirring force. In the {\sc lab} frame, this perturbation
is described by adding to the  Hamiltonian the following term:
\begin{equation}
U_{LAB}(t) = \int d{\bf r}\: \delta v_{LAB}({\bf r},t) \rho({\bf r}).
\end{equation}
In the previous expression, we have neglected any dependence 
on the axial coordinate. Furthermore, in order
to excite predominantly surface modes, a multipole expansion of 
the external potential $\delta v_{LAB}(r, \theta,t)$ should not contain high multipole terms.
This means that it will be a smooth function of $r$ near the boundary of the boson droplet, i.e. for 
$r \approx R$. If we further assume that the external potential contains no monopole mode, i.e.
\begin{equation}
\int^{2\pi}_{0} d\theta \: \delta v_{LAB}(r, \theta,t ) = 0,
\end{equation}
then we can write 
\begin{equation}
U_{LAB}(t) =  \int d{\bf r}\: \delta v_{LAB}({\bf r},t) \left[ \rho({\bf r}) - \rho_{\rm o}(r)\right]
\approx \int^{2\pi}_{0} d\theta \: \delta v_{LAB}(R, \theta,t)\:  \int^{+\infty}_{0} dr \: 
r \: \left[ \rho(r, \theta) -\rho_{\rm o}(r) \right],
\end{equation}
where the last expression follows from the smoothness of the potential near the boundary of the
droplet. The expression integrated over $r$ is the boundary density operator, $j(\theta)$.
Hence
\begin{equation}\label{pert}
U_{LAB}(t) = \int^{2\pi}_{0} d\theta \: \delta v_{LAB}(R, \theta,t)\: j(\theta). 
\end{equation}
This corresponds to the perturbation discussed in the main text, but expressed in the {\sc lab} frame.
The {\sc rot} and {\sc lab} are related by the unitary transformation
\begin{equation}\label{unit}
{\cal U}(t) = e^{iL \Omega t/\hbar}.
\end{equation}
Notice that this transformation will affect only those correlation functions which are time-dependent (e.g.
like the dynamic structure factor). Time-independent correlation functions, such like the one-body
density matrix and its diagonal part, the equilibrium density profile, 
are unchanged when going from one reference frame to the other. 
To see how the dynamic structure factor computed in the rotating  ({\sc rot}, for short) frame is
related to measurements in the {\sc lab} frame, which is where the measurement apparatus
is actually placed, we need to work out the effects of the perturbation, Eq.~(\ref{pert}), in
{\sc lab}.  Starting from the time-dependent Schr\"odinger equation,
\begin{equation}\label{schro}
i\hbar \partial_t |\Phi_{LAB}(t) \rangle = \left[ H_{LAB} + U_{LAB}(t) \right] | \Phi_{LAB}(t) \rangle.
\end{equation}
we find that, to  linear order in $U_{LAB}(t)$, the wave function is modified to
\begin{equation}
| \Phi^{I}_{LAB}(t) \rangle \approx | \Phi^{I}_{LAB}(-\frac{T_{\rm o}}{2}) \rangle - 
\frac{i}{\hbar} \int_{-\frac{T_{\rm o}}{2}}^{t}
dt' \: U^{I}_{LAB}(t') |\Phi^{I}_{LAB}(t') \rangle,
\end{equation}
where the interaction representation has been used:
\begin{eqnarray}
| \Phi^{I}(t) \rangle &=& e^{iH_{LAB}t/\hbar} | \Phi(t) \rangle,\\
U^{I}(t) &=& e^{iH_{LAB}t/\hbar} U(t) e^{-iH_{LAB}t/\hbar}.
\end{eqnarray}
Before going any further, it is convenient to recall that
the Hamiltonians in the {\sc rot} and {\sc lab} frames are related by the following
expression
\begin{equation}
H_{ROT} = H_{LAB} - \Omega L, 
\end{equation}
where $\left[L, H_{LAB} \right] = 0$, as the trap potential is cylindrically symmetric (this is not so after
$U_{LAB}(t)$ is turned on). This expression is obtained by transforming the time-dependent
Schr\"odinger equation, (\ref{schro}) to the   {\sc rot}  frame using Eq.~(\ref{unit}). Indeed, it is
the quantum analog of the classical relationship $E_{ROT} = E_{LAB} - \Omega L$.

Assuming that, in the {\sc rot} frame, at time $t = -\frac{T_{\rm o}}{2}$ the system is  in a state
$|\Phi_n \rangle$ with probability given by the Boltzmann weight $e^{-E^{ROT}_n/T}$,
we  compute the probability per unit time for large $T_{\rm o}$ that the system 
is found, also in the {\sc rot} frame, in a state $|\Phi_m\rangle$ at $t = T_{\rm o}/2$ 
(we choose $H_{ROT} |\Phi_{\alpha}\rangle = E^{ROT}_\alpha |\Phi_\alpha \rangle$ and 
$L  |\Phi_\alpha\rangle = L_\alpha  |\Phi_\alpha\rangle $, $\alpha = n, m$. The calculation, however, 
is performed in the {\sc lab} frame). The result
can be written as 
\begin{equation}\label{rate}
\Gamma(T,\Omega) = \frac{1}{2\pi} \sum_{l=-\infty}^{+\infty} \int\limits_{-\infty}^{+\infty} 
\frac{d\omega}{2\pi}\,  \left|\delta v_{LAB}(R,l,\omega) \right|^2 \:
S_{ROT}(l,\omega - l \Omega).
\end{equation}
Therefore, we can define the dynamic structure factor in the {\sc lab} frame as
\begin{equation}\label{transf}
S_{LAB}(l,\omega) = S_{ROT}(l,\omega - l \Omega).
\end{equation}
where the dynamic structure  factor,
\begin{equation}
S_{ROT}(l,\omega) = \int_{-\infty}^{+\infty} dt \:  \int^{2\pi}_{0} d\theta' \: e^{i\omega t-il\theta'}
\langle j(\theta',t) j(0) \rangle = \frac{2\pi\hbar}{Z}
\sum_{\alpha, \beta} e^{-E^{ROT}_\alpha/T} |\langle \beta | j(l) | \alpha \rangle |^2 
\: \delta(\hbar \omega + E^{ROT}_{\alpha} - E^{ROT}_{\beta}).  
\end{equation}
There is a simple explanation for the transformation law, Eq.~(\ref{transf}). The shift $\omega \to
\omega - \Omega l$ is precisely the frequency in the {\sc rot} frame of an excitation 
of angular momentum $\hbar l$ and energy $\hbar \omega$ in the {\sc lab} frame. The other
component in the dynamic structure factor are the oscillator strengths 
$ |\langle \alpha | j(l) | \beta \rangle |^2 $ for transitions induced between different
eigenstates of $H_{ROT}$ (and also of $H_{LAB}$, since it commutes with $L$).
Since the {\sc rot} and {\sc lab} frames are related by a unitary transformation,
which leaves these probabilities unchanged, only the delta function in $\omega$
is affected by the change of reference frame. 

  Eq. ~(\ref{rate}) implies that we can measure $S_{LAB}(l,\omega)$ by cleverly choosing
the external perturbation $\delta v_{LAB}$. It is  important to point out 
that in the discussion in the main text, we have omitted
any reference to the dependence of the intensity of the response on the droplet radius,
$R$, coming from $\delta v_{LAB}(R, l, \omega)$. When analyzing
the experimental data this fact must be taken into account since $R$, defined as the 
semiclassical radius of the droplet, depends on the effective filling fraction $\nu$. For
instance, for a Laughlin droplet we found that $R = \sqrt{m N} \: \ell$, and in general
$R = \sqrt{N/\nu}\: \ell$, being the {\it average} density given by $\nu/\pi\ell^2$.
Thus, besides the factor $\nu = p/(p+1)$, with $p$ a non-negative integer, discussed 
in previous sections, the strength of the peak in response to an $l$-polar 
potential $\delta v_{LAB} \sim e^{il \theta}$ is also proportional to $R^l$.
This factor is especially important when analyzing the dipole mode  (i.e. $l = 1$).  This is because the
energy and the peak intensity  of this mode are given, for a many-body system confined in a harmonic 
trap, by the generalized Kohn's theorem~\cite{Yip91,Lipparini97}. This theorem follows from the
fact in a harmonic trap the motion of the center of mass decouples from  
other motions of the system. Introducing the operators:
\begin{eqnarray}
\mathbf{\Pi} &=& \sum_{j=1}^{N} \left[{\bf p}_j - M\Omega \hat{\bf z}  \times {\bf r}_j \right],\\
\mathbf{R} &=& \frac{1}{N}\sum_{j=1}^{N} {\bf r}_j,
\end{eqnarray}
the Hamiltonian (in the {\sc rot} frame) splits into two commuting parts (cf. Eq.~(\ref{eq2})),
\begin{eqnarray}
H_{ROT} &=& H_{CM} + H_{R},\\
H_{CM} &=& \frac{\mathbf{\Pi}^2}{2MN} + \frac{MN}{2} (\omega^{2}_{\perp} - \Omega^2) {\bf R}^2.
\end{eqnarray}
To study the excitation of the center-of-mass degrees of freedom it is convenient to 
use the set {\it independent} operators,
\begin{eqnarray}
A^{\dag} &=& \frac{\Pi_x + i \Pi_y}{M} - i (\omega_{\perp} +\Omega) N (X + i Y), \\
B^{\dag} &=& \frac{\Pi_x - i \Pi_y}{M}  +   i (\omega_{\perp} -\Omega) N (X - i Y),
\end{eqnarray}
which have the following properties: 
\begin{eqnarray}
\left[H_{ROT},A^{\dag}\right] &=& \hbar \omega_A A^{\dag},\\ 
\left[H_{ROT},B^{\dag}\right] &=& \hbar \omega_B B^{\dag},\\
\left[L, A^{\dag}\right] &=& \hbar A^{\dag},\\
\left[L, B^{\dag}\right] &=& -\hbar B^{\dag},\label{reduc}
\end{eqnarray}
where $\omega_A = \omega_{\perp} - \Omega$, i.e. the energy of $l=1$ surface mode, 
which in the {\sc lab} frame will have an energy equal to $\hbar \omega_{\perp}$.   
The theorem also predicts the existence of another Kohn mode,
at $\omega_B = \omega_{\perp} + \Omega$ in the {\sc rot}
frame. Notice that although this mode has much higher energy in the
{\sc rot} frame, it actually reduces the angular momentum by one quantum
and therefore, application of the formula $E_{ROT} = E_{LAB} - \Omega L$,
yields an excitation energy in the {\sc lab} frame equal to $\omega_{\perp}$. 
Thus we conclude that the center-of-mass modes created by acting with
the operators $A^{\dag}$ and $B^{\dag}$ on the many-body ground state
are {\it degenerate} in the {\sc lab} frame. However, since the system
is in thermal equilibrium in the {\sc rot} frame, it is the energy in this system
that enters in the Boltzmann factor, $e^{-E^{ROT}_{n}/T}$. Hence, at 
temperatures $T \ll \hbar \Omega$, the probability of finding the system 
in the state $A^{\dag}|0\rangle$ 
is overwhelmingly larger than the probability of  finding it in $B^{\dag}|0\rangle$.
By their quantum numbers we can identify $A^{\dagger} \sim s_1 = \sum_{j=1}^{N} z_j$,
and $B^{\dagger} \sim d_1 \sim \sum_{j=1}^{N} \bar{\pi}_j \sim \sum_{j=1}^{N} \bar{z}_j$, and
as we argued in Sect.~\ref{sectlaughlin}, the former yields a low-lying state whereas the
latter does not.

 Let us finally compute the the  peak intensity of the mode  created by $A^{\dag}$: 
\begin{equation}
f(l=1, T=0) = |\langle 1  | N(X+ iY) | 0 \rangle |^2 = \frac{1}{4\omega^2_{\perp}} 
|\langle 1  | (A^{\dagger} - B) | 0 \rangle |^2,
\end{equation}
where $|1\rangle = A^{\dag}|0\rangle/\sqrt{|[A,A^{\dag}]|}$.  Thus we 
obtain
\begin{equation}
f(l=1,T=0) = \frac{\hbar N}{M\omega_{\perp}} = N\ell^2. 
\end{equation}
This is in agreement with the predictions of effective field theory once the 
dependence on $R$ has been taken into account. For instance, consider a
Laughlin droplet perturbed by the dipolar potential
$ N(X+iY) e^{-i\omega t}$, i.e. $\delta v_{LAB}(R,\theta, t) \sim R e^{i\theta}e^{-i\omega t}$. 
Using Eq.~(\ref{rate}) and the results of Sect.~\ref{sectlaughlin} 
\begin{equation}
f(l=1,T=0) = R^2 \frac{1}{m} = N \ell^2 
\end{equation}
where we have used $R =\sqrt{mN} \ell$. This result is identical to the one
obtained by the generalized Kohn's theorem, as it should be. For
states with several branches of surface phonons $R^2 = N\ell^2/\nu$, and 
since the total spectral weight of $S_{LAB}(l,\omega)$ for $l=1$ equals
$\nu = p/(p+1)$ ($p$ integer), and  taking into account that $R^2 \nu =  N \ell^2$, 
we see that all the spectral weight must be concentrated in  a single peak
at $\omega = \omega_{\perp}$.
Therefore, all corrections to the energy of the surface modes must
vanish for $l = 1$, as discussed in Sect.~\ref{other},  making 
all branches degenerate at this particular angular momentum. 

\section{Plasma formalism and Vertex Operators}\label{appb}

 In this appendix we shall introduce the plasma formalism, which is a convenient
tool to study some properties of quantum Hall wave functions. However,
our main goal here is to establish a deep connection 
between this formalism and the correlation functions of a two-dimensional
gaussian field theory~\cite{Wen95,Moore91}. As we show below for the Laughlin state, the $N$-particle
wave function can be obtained as the multipoint correlation function 
of $N$ objects, known as vertex operators, plus another operator, which yields the
gaussians  omitted so far. The vertex operators 
are then related to  the boundary field operators  introduced in
Sect.~\ref{sectlaughlin}. The construction  shows explicitly the relationship
between the low-energy description of  surface modes, and
the bulk wave functions. The results
of this section are used in Sect.~\ref{vecbosons}
for the slightly more complicated case of vector bosons. Finally, we also show
how to obtain the one-body density matrix, Eq.~(\ref{dmt0}).

Consider the wave function for Laughlin state
\begin{equation}
\Phi_{m}(z_1, \ldots, z_N) = {\cal N} \, 
\prod_{i<j=1}^{N} (z_i - z_j)^{m} \prod_{i=1}^{N} e^{-|z_i|^{2}/2\ell^2}.
\end{equation}
The (unnormalized) probability density can be written as 
\begin{equation}\label{prob}
P_m({\bf r}_1, \ldots, {\bf r}_N)  = |\Phi_m( z_1, \ldots, z_N) |^2 = 
\exp \left[ -\beta U({\bf r}_1, \ldots, {\bf r}_N) \right],
\end{equation}
where $\beta = 2m$ and  
\begin{equation}\label{plasmapot}
U({\bf r}_1, \ldots, {\bf r}_N) = \frac{1}{2m\ell^2}\sum_{i=1}^{N} |z_i|^2 - \sum_{i<j = 1}^{N}
 \log |z_i-z_j|, 
\end{equation}
is the potential energy of a one-component classical  plasma  in two dimensions.
The plasma consists of  $N$ identical particles with unit charge moving on a 
uniform neutralizing background of charge density $\rho_{\rm o} = 1/m\pi \ell^2$.
The plasma is neutral if the background takes up a disk of radius $R = \sqrt{(\pi 
\rho_{\rm o})^{-1} N} = \sqrt{m  N} \ell$. To understand 
the analogy, let us recall that in two dimensions the Coulomb potential due to a unit
charge reads
\begin{equation}
g({\bf r} - {\bf r}') = -\log |{\bf r} - {\bf r}' | + {\rm const.} = -\log |z - z'|  + {\rm const.}
\end{equation}
In addition,  a uniform background of charge $\rho_{\rm o}$ creates a potential at $\bf r$
(with $|{\bf r}| < R$) given by
\begin{equation}
g_{\rm BG}({\bf r}) = \int_{|{\bf r}'| < R} d{\bf r}' \, g({\bf r}-{\bf r}')\: \rho_{\rm o} = -\frac{\pi \rho_{\rm o}}{2} |z|^2,
\end{equation}
as can be easily checked using Poisson's equation: $\nabla^2 g_{\rm BG}({\bf r}) = -2 \pi \rho_{\rm o}$.
Putting everything together, Eq.~(\ref{plasmapot}) is obtained.   

 It will be useful to rewrite~(\ref{prob}) as follows
\begin{equation}
P_{m}[\rho] = \frac{Z_{\rm gauss}[\rho]}{Z_{\rm gauss}[0]},
\end{equation}
where $Z_{\rm gauss}[\rho]$ is the functional integral
\begin{equation}\label{gauss}
Z_{\rm gauss}[\rho] = \int \left[ du({\bf r}) \right] \, e^{- \beta \int d{\bf r}\, 
\left[ \frac{1}{4\pi}\left(\nabla u({\bf r}) \right)^2 - i \rho({\bf r}) u({\bf r}) \right] },
\end{equation}
where $\rho({\bf r}) = -\sum_{i=1}^{N} \delta({\bf r} - {\bf r}_i) + \rho_{\rm o} \vartheta( R - |{\bf r}|)$ is 
the total charge density of the plasma ($\vartheta(r)$ is the step function). The above expression
can be shown to be equivalent to Eq.~(\ref{prob}) by 
shifting
\begin{equation}
u({\bf r}) = u'({\bf r}) + i  \int d{\bf r}' \: g({\bf r}- {\bf r}') \rho({\bf r}'),
\end{equation}
which cancels the term linear in $u({\bf r})$ in~(\ref{gauss}) and yields the
following expression for the potential energy of the plasma 
\begin{equation}
U({\bf r}_1, \ldots, {\bf r}_N) = U[\rho] = \frac{1}{2} \int d{\bf r} \, \rho({\bf r}) g({\bf r}-{\bf r}') \rho({\bf r}').
\end{equation}
Substituting $\rho({\bf r})$ defined as above, this expression reduces to~(\ref{plasmapot})
(the charge self-energies can be removed by redefining 
$g({\bf r}) = -\log\left[ (|{\bf r}|^2 + \alpha^2)/\alpha^2\right]$ and 
carefully taking $\alpha \to 0$ at the end).

  We have taken so much work to show that $P({\bf r}_1, \ldots, {\bf r}_N)$ can be written
as the functional integral~(\ref{gauss}) because this form allows us to make a very
useful connection. Indeed Eq.~(\ref{gauss}) is the ``generating functional'' of a two-dimensional 
field theory: For arbitrary $\rho({\bf r})$, it generates all correlation functions of a gaussian
field theory, which describes, e.g., the classical statistical mechanics of a two-dimensional 
superfluid or the quantum mechanics in imaginary time of a one-dimensional 
quantum fluid~\cite{Tsvelik95}. If we specialize to $\rho({\bf r}) = 
\rho_{\bf o} \vartheta(R - |{\bf r}|) - \sum_{i=1}^{N} \delta({\bf r}- {\bf r}_i)$, then Eq.~(\ref{gauss})
generates $N$-point correlation functions of the form
\begin{equation}
P({\bf r}_1, \ldots, {\bf r}_N) = \langle e^{-2imu({\bf r}_1)} \ldots e^{-2imu({\bf r}_N)} 
e^{+2im  \int\limits_{|{\bf r}| < R} d{\bf r} \,   u({\bf r}) \rho_{\rm o} } \rangle,
\end{equation}
If instead of regarding this correlation function as a statistical average in two dimensions,
we think of it as a quantum expectation value in one dimension, the objects $e^{2mi u({\bf r})}$ represent
operators, termed ``vertex'' operators. The field operator $u({\bf r})$ obeys
the equation of motion $\nabla^2 u({\bf r}) = 0$, i.e. the Laplace equation. This also corresponds to the equation
of motion of a phonon field in imaginary time. Recalling that any solution of Laplace's 
equation can be written as the linear combination of an analytic and an anti-analytic 
part, let us write 
\begin{equation}
u({\bf r}) = \frac{1}{2\sqrt{m}}\left[\bar{\varphi}(\bar{z}) - \varphi(z) \right], 
\end{equation}
where $z = x + i y$ and $\bar{z} = x - i y$ and the normalization $1/\sqrt{m}$ has been
introduced for later convenience. Next, notice that the above field operators
obey the (first order) equations of motion:
\begin{eqnarray}
\partial_z  \bar{\varphi}(\bar{z}) \equiv \frac{1}{2}\left(\partial_x - i\partial_y \right) \bar{\varphi}( x - i y) &=& 0, \\
\partial_{\bar{z}}\varphi(z) \equiv  \frac{1}{2} \left( \partial_x + i \partial_y \right) \varphi( x + iy) &=& 0.
\end{eqnarray}
The structure of these equations  resembles very much the equation
of motion for the {\it chiral} phonon field $\phi(\theta)$ introduced in Sect.~\ref{sectlaughlin}.
Recall that there it was found that $\phi(\theta,t) = \phi(\theta - \upsilon t)$, which 
in imaginary time $\tau = it$ obeys $\left(\upsilon^{-1} \partial_{\tau} + 
i \partial_\theta  \right) \phi(\upsilon \tau - i\theta) = 0$. In an analogous way as we did for $\phi(\theta)$ 
in terms of $e^{i\theta}$,
we can expand the fields  $\varphi$ and $\bar{\varphi}$ in a series of $z$ and $\bar z$, 
respectively. These expansions read:
\begin{eqnarray}
\varphi(z) &=& -\varphi_{\rm o} - i q \log z 
+ \sum_{l>0} \frac{1}{\sqrt{l}} \left[z^{-l} c(l) + z^{l} c^{\dagger}(l)\right], \\
\bar{\varphi}(\bar{z}) &=&\, \, \, \bar{\varphi}_{\rm o} + 
i \bar{q} \log \bar{z} + \sum_{l>0} \frac{1}{\sqrt{l}} \left[z^{-l} \bar{c}(l) 
+ \bar{z}^{l} \bar{c}^{\dagger}(l)\right],
\label{modesbarphi}
\end{eqnarray}
where $\left[q, \varphi_{\rm o} \right] = i, \left[c(l), c^{\dagger}(l') \right] = \delta_{l,l'}$, 
and similar expressions for $\bar{q}, \bar{ \varphi_{\rm o} }$ and $\bar{c}(l), \bar{c}^{\dagger}(l)$.
Returning to the vertex operators, which are our main interest in this appendix, we notice that,
in operator formalism, it is necessary to normal- order them as follows:
\begin{equation}
V_{\beta}(z) = \, : e^{i\beta \varphi(z)} : \, \, = e^{-i\beta \varphi_{\rm o}}\: e^{\beta q \log z}\:  
e^{i \beta \sum_{l>0} z^{l} c^{\dagger}(l)/\sqrt{l} } \: e^{i \beta \sum_{l>0} z^{-l} c(l)/\sqrt{l}},
\end{equation}
and a similar expression holds for the vertex operators of $\bar{\varphi}(\bar{z})$. This is
needed in order to avoid divergences, which are related 
to the ``charge self-energy'' problem encountered above
when using the functional integral approach. With these provisos, 
the probability distribution of the Laughlin state can be written
as:
\begin{equation}
P_m({\bf r}_1,\ldots, {\bf r}_N) = \bar{\Phi}_m(\bar{z}_1,\ldots, \bar{z}_N) \Phi_m(z_1, \ldots, z_N), 
\end{equation}
where
\begin{eqnarray}
\bar{\Phi}_m(\bar{z}_1,\ldots, \bar{z}_N) &=& \langle \bar{V}_{-\sqrt{m}}(\bar{z}_1) \ldots 
\bar{V}_{-\sqrt{m}}(\bar{z}_N) \, :e^{+i\sqrt{m} \int\limits_{|z| < R} d^{2}z\, 
 \bar{\varphi}(\bar{z}) \rho_{\rm o}} : \rangle,\\
\Phi_m(z_1,\ldots, z_N) &=& \langle V_{+\sqrt{m}}(z_1)\ldots 
V_{+\sqrt{m}}(z_N) \, :e^{-i\sqrt{m}\int\limits_{|z|<R} d^2z\,  \varphi(z) \rho_{\rm o}}:  \rangle.
\end{eqnarray}
 Therefore, the probability distribution
for the Laughlin state splits into the product of two correlation functions of the gaussian
theory. Let us now demonstrate that each of these  equals either the Laughlin state or
its anti-analytic version (i.e. its complex conjugate). 
To this end, it is useful to further investigate some of
the properties of  the vertex operators. For example, their two-point correlation function reads:
\begin{equation}\label{twop}
\langle V_{\beta}(z) V_{-\beta}(z') \rangle = (z-z')^{-\beta^2},
\end{equation}
as it can be shown from the identity $e^{A} e^{B} = e^{C} e^{B} e^{A}$ for any 
two operators with a c-number commutator $C = [A, B]$. The generalization
to multi-point correlation functions can be obtained in a similar fashion:
\begin{equation}
\langle V_{\beta_1}(z_1) \ldots V_{\beta_N}(z_N) \rangle = \prod_{i<j} \left(z_i-z_j\right)^{\beta_i\beta_j}
\end{equation}
provided that $\sum_{i=1}^{N} \beta_i = 0$. Using this result, it can be shown that 
\begin{equation}
\Phi_m(z_1, \ldots, z_N) = e^{iw(z_1,\ldots,z_N)} \: \prod_{i<j}(z_i- z_j)^{m} 
\prod_{i=1}^{N} e^{-|z_i|^{2}/2\ell^2},
\end{equation}
i.e. the Laughlin state. The phase factor $e^{iw(z_1,\ldots, z_N)}$ cancels out after taking the product
with the anti-analytic part $\bar{p}(\bar{z}_1, \ldots )$, and we shall not consider 
it any further. 

 To end this discussion, let us make  
 firmer the connection  
to the low-energy description of surface modes, which we have loosely established above. 
This can be achieved by a  re-examination of the field operator found in 
Sect.~\ref{sectlaughlin}. There we showed that the boson field operator is related to
the operator $e^{\pm i m\phi(\theta)}$, which 
can be shown to be proportional to a vertex operator. 
Indeed, if we normal order it properly, we obtain ($\alpha$ is a short distance cut-off):
\begin{equation}
e^{\pm i m\phi(\theta)}  = \alpha^{m/2} e^{-im\theta/2} \: \, : e^{\pm i m\phi(\theta)} : \, \, .
\end{equation}
We next notice that by comparing the 
mode expansions, Eq~(\ref{modesbarphi}) 
and Eq.~(\ref{modes}),  one can
identify  $\phi(\theta) = \bar{\varphi}(\bar{z} = e^{-i\theta})/\sqrt{m}$.
Thus we arrive at
\begin{eqnarray}
\label{fop1}
\Psi(\theta) &=& A \,  e^{+im (N-1) \theta} \, \bar{V}_{+\sqrt{m}}(\bar{z} = e^{-i\theta}), \\
\Psi^{\dagger}(\theta) &=& A \,  e^{imN\theta} \bar{V}_{-\sqrt{m}}(\bar{z} = e^{-i\theta}). \label{fop2},
\end{eqnarray}
from which one can readily show that the one-body density matrix at the boundary,
\begin{equation}
G(\theta-\theta') = \langle \Psi^{\dagger}(\theta) \Psi(\theta') \rangle
=  {\rm const} \times e^{-iml_{\rm o}(\theta -\theta')}  \left[ \sin \left(\frac{(\theta-\theta')}{2} \right) \right]^{-m}, 
\end{equation}
as follows from Eqs.~(\ref{fop1},\ref{fop2}), 
and~(\ref{twop}). Therefore, if we define the boson operator in the LLL as
$\Psi_{B}^{\dagger}(\bar{z}) \equiv \bar{V}_{-\sqrt{m}}(\bar{z})$, the Laughlin
wave function can be written in the very appealing way,
\begin{equation}
\bar{\Phi}_m(\bar{z}_1, \ldots, \bar{z}_N) = \langle \Psi_{B}^{\dagger}(\bar{z}_1) \ldots
\Psi_{B}^{\dagger}(\bar{z}_N) \, :e^{+i\sqrt{m} \int\limits_{|z| < R} d^{2}z\,  \bar{\varphi}(\bar{z}) \rho_{\rm o}}: \rangle.
\end{equation}
This is the main result of this appendix.


\begin{thebibliography}{30}
\bibitem{trap}
In this work we shall assume that the trapping potential may 
be anisotropic, with $\omega_{||} \geq \omega_{\perp}$, where
$\omega_{||}$ is the trap frequency along the rotation axis, Z, 
and $\omega_{\perp}$ the trap frequency perpendicular to Z.
\end{thebibliography}
\end{document}